\documentclass[a4paper,10pt]{article} 				
\usepackage{amsmath} 								
\usepackage{amsfonts}
\usepackage{amssymb}
\usepackage{hyperref}
\hypersetup{colorlinks = true, linkcolor = red, citecolor = blue}

\usepackage{color}
\definecolor{prettyblue}{rgb}{0.20,0.40,0.65}

\usepackage{mathtools}
\usepackage{cleveref}
\usepackage{graphicx}
\usepackage[font=small,labelfont=bf]{caption}
\usepackage[T1]{fontenc}
\usepackage{authblk}								
\usepackage[english]{babel}	               					
\usepackage{csquotes}
\usepackage[round]{natbib}							
\usepackage[titletoc,toc,title]{appendix}			

\newcommand{\qed}{\nobreak \ifvmode \relax \else
      \ifdim\lastskip<1.5em \hskip-\lastskip
      \hskip1.5em plus0em minus0.5em \fi \nobreak
      \vrule height0.75em width0.5em depth0.25em\fi}
      
\usepackage{setspace}
\title{Computing Integrated Information}
\author[1]{Stephan Krohn}
\author[1,2]{Dirk Ostwald\small}\date{}
\affil[1]{Computational Cognitive Neuroscience Laboratory \authorcr Department of Education and Psychology, Habelschwerdter Allee 45, 14195 Berlin, Freie Universit\"at Berlin}
\affil[2]{Center for Adaptive Rationality \authorcr Max-Planck Institute for Human Development, Berlin}

\begin{document}
\maketitle

\vspace{-2mm}
\begin{center} stephan.krohn@fu-berlin.de \end{center}
\vspace{2mm}


\begin{abstract}

\noindent \textcolor{black}{Integrated information theory (IIT) has established itself as one of the leading theories for the study of consciousness. IIT essentially proposes that quantitative consciousness is identical to maximally integrated conceptual information, quantified by a measure called $\Phi^{max}$, and that phenomenological experience corresponds to the associated set of maximally irreducible cause-effect repertoires of a physical system being in a certain state. However, in order to ultimately apply the theory to experimental data, a sufficiently general formulation is needed. With the current work, we provide this general formulation, which comprehensively and parsimoniously expresses $\Phi^{max}$ in the language of probabilistic models. Here, the stochastic process describing a system under scrutiny corresponds to a first-order time-invariant Markov process, and all necessary mathematical operations for the definition of $\Phi^{max}$ are fully specified by a system's joint probability distribution over two adjacent points in discrete time. We present a detailed constructive rule for the decomposition of a system into two disjoint subsystems based on flexible marginalization and factorization of this joint distribution. Furthermore, we suspend the approach of interventional calculus based on system perturbations, which allows us to omit undefined conditional distributions and virtualization. 
We validate our formulation in a previously established discrete example system, in which we furthermore address the previously unexplored theoretical issue of quale underdetermination due to non-uniqueness of maximally irreducible cause-effect repertoires, which in turn also entails the sensitivity of $\Phi^{max}$ to the shape of the conceptual structure in qualia space. 
In constructive spirit, we propose several modifications of the framework in order to address some of these issues.} 

\end{abstract}
\section{Introduction}\label{sec:Introduction}

\textcolor{black}{Integrated information theory \citep{Tononi2004, Tononi2005, Tononi2008, Tononi2012,Oizumi2014,Tononi2015,Tononi2016} has established itself as one of the most prominent theories in the study of the physical substrates of consciousness. Integrated information theory (IIT) essentially proposes that quantitative consciousness, i.e. the degree to which a physical system is conscious, is identical to its state-dependent level of maximally integrated information, which can be quantified in a measure called ``$\Phi^{max}$''.}  Integration here means that the information generated by the system as a whole is in some measurable sense more than the information generated by its parts and intuitively corresponds to finding an index of a system state's functional irreducibility.
\textcolor{black}{Intriguingly, IIT also equates the set of maximally integrated cause-effect repertoires associated with a system state to qualitative consciousness, i.e. the actual phenomenological experience or "what-it-is-like-ness" (\citep{Nagel1974}) of a physical system being in a certain state, and thus aims at nothing less than a formal description of a quale.
 While this approach is not undisputed \citep[e.g.,][]{Aaronson2014, Cerullo2015}, IIT has both explanatory and predictive power and thus the idea of measuring integrated information has by now gained widespread popularity in the cognitive neuroscience literature and beyond \citep[e.g.,][]{Balduzzi2008,Balduzzi2009,  Deco2015,Koch2016,Tegmark2016}.}
\textcolor{black}{
Ultimately, the aim of integrated information theory must be to evaluate $\Phi^{max}$ as a theoretically derived measure of quantitative consciousness based on empirical data, such as electroencephalographic (EEG) or functional magnetic resonance imaging (fMRI) recordings. However, to date, the theory underlying the integrated information measure is primarily developed with respect to specific examples, usually low-dimensional discrete state systems that implement logical operations. For the computation of $\Phi^{max}$ for a wide variety of systems and based on different data types, a sufficiently general formulation of IIT is required.  
With the current work, we provide this general formulation, starting from the most recent instantiation of the theory called "III 3.0" \citep{Oizumi2014}, which features several important theoretical advances over previous versions of IIT. Henceforth, we thus use the abbreviation ``IIT'' to refer exclusively to integrated information theory 3.0 as developed by \citet{Oizumi2014}, unless explicitly stated otherwise.
In the Methods section, we first present a comprehensive formulation of IIT with respect to the general language of probabilistic models, by which we simply mean joint probability distributions over random entities \citep[e.g.,][]{Efron2016,Gelman2014, Barber2012, Murphy2012}. We derive a constructive rule for the decomposition of a system into two disjoint subsets, central to the definition of information integration. Moreover, we show that our general formulation improves parsimony, as it suspends interventional calculus, reduces virtualization to distribution factorization, and eschews undefined conditional probability distributions. In contrast, all mathematical operations presented herein are sufficiently specified by a system's joint probability distribution over two adjacent points in time by flexible marginalization and factorization.
We then validate our general formulation in the Results section by evaluating $\Phi^{max}$ in a previously established discrete state example system. Here, we also illustrate the theoretical issue of "quale underdetermination", and we show that the current definition of $\Phi$ combines IIT's quantitative and qualitative measures of consciousness, which we suggest be better disentangled. Finally, we relate our approach for computing $\Phi$ to similar endeavors in the literature, discuss some open questions in IIT and propose constructive modifications of the framework to overcome some of the above-mentioned issues.}

\subsection{Notation, Terminology, and  Implementation}\label{sec:Notation}

A few remarks on our notation of probabilistic concepts are in order. To denote random variables/vectors and their probability distributions, we use an applied notation throughout. This means that we eschew a discussion of a random entity's underlying measure-theoretic probability space model  \citep[e.g.,][]{Billingsley2008}, and focus on the random entity's outcome space and probability distribution. For a random variable/vector $X$, we denote its distribution by $p(X)$, implicitly assuming that this may be represented either by a probability mass or a probability density function. To denote different distributions of the same random variable/vector, we employ subscripts. For example, $p_a(X)$ is to indicate a probability distribution of $X$ that is different from another probability distribution $p_b(X)$. In the development of integrated information, stochastic (conditional) dependencies between random variables are central. To this end, we use the common notation that the statement $p(X|Y) = p(X)$ is meant to indicate the stochastic independence of $X$ from $Y$ and the statement $p(X|Y,Z) = p(X|Z)$ is meant to indicate the (stochastic) conditional independence of $X$ on $Y$ given $Z$ \citep{Dawid1979,Geiger1990}. Since the notion of a system subset being in a particular state is crucial for the definition of $\Phi$, we refer to a given subset by the superscript $S$ and the realization of a state with an elevated asterisk.\\ 
\textcolor{black}{Since IIT comes with its own terminology, it may be helpful to highlight some expressions used throughout the manuscript. In the following, by "system" we interchangeably mean a network of physical elements described by a corresponding set of random entities. A "purview" refers to the notion of considering a particular subset of random entities in describing the system. For any subset being in a particular state at a specific time, the "cause repertoire" of that state refers to a conditional probability distribution over past states, and the "effect repertoire" describes the conditional distribution over future states. A "partition" means rendering the system into two independent parts. The terms "concept" and "conceptual structure" refer to maximally integrated cause and effect repertoires and are explained in the context of our formulation in \cref{sec:comp_exclu}.    
The reader wishing to retrace our formulation of IIT will find all Matlab code (The MathWorks, Inc., Natick, MA, United States) developed for the implementation of the below and the generation of the technical figures herein from the corresponding author and the Open Science Framework (\url{https://osf.io/nqqzg/}).}

\section{\textcolor{black}{Methods: Defining $\Phi$}}\label{sec:Theory}
\subsection{System model}

\begin{figure}[htp!]
\centering
\includegraphics[width = 85 mm]{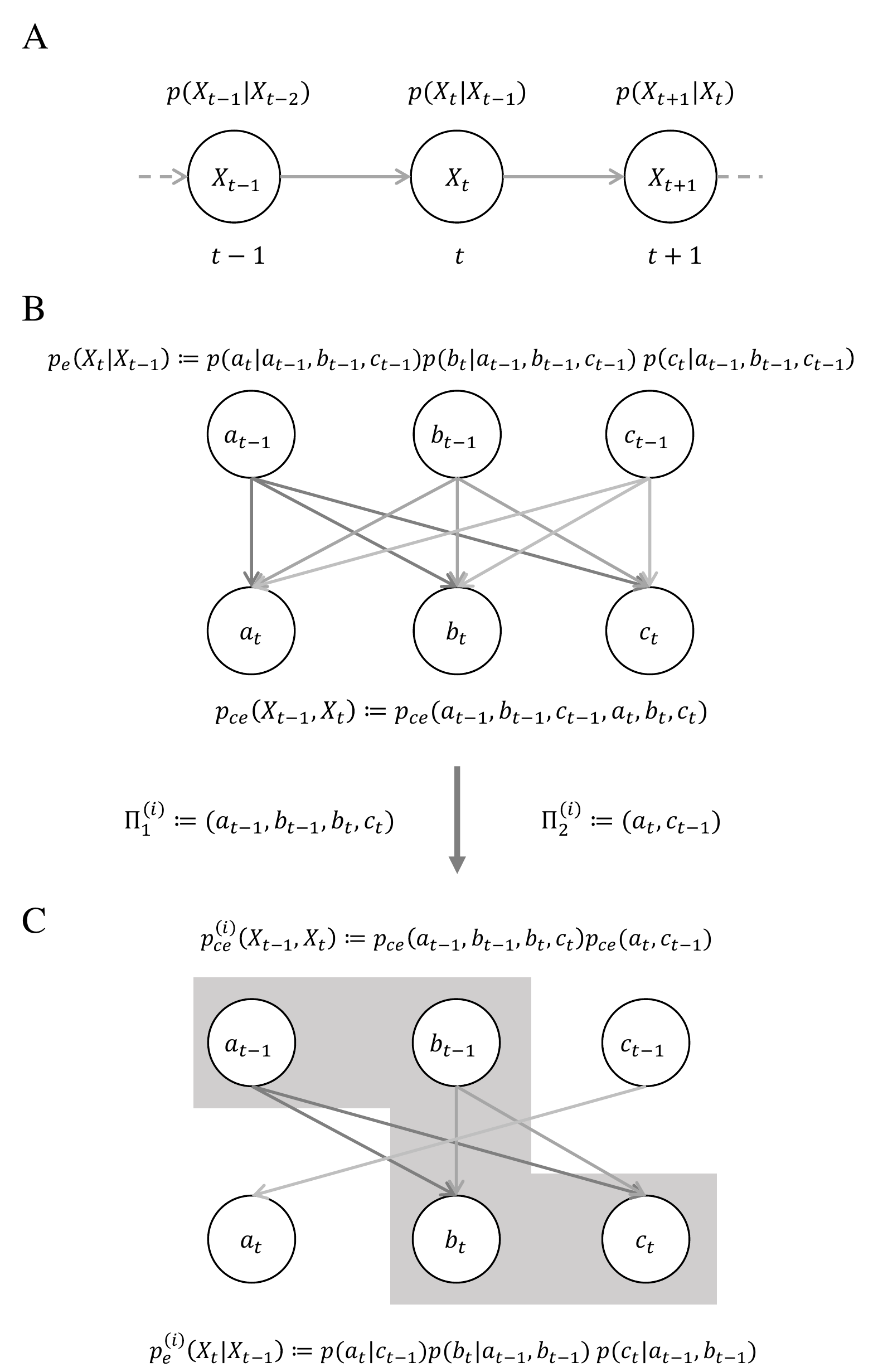}
 \caption{\textbf{System model and system decomposition for integrated cause-effect information}. IIT models the system of interest by a time-invariant first-order Markov process, depicted in Panel A as graphical model \citep[e.g.,][]{Bishop2006}. Nodes denote random vectors and directed links denote the stochastic dependence of the child node on the parent node. Panels B and C display the exemplary decomposition of a three-dimensional system with state vector $X_t := (a_t, b_t, c_t)$ as a graphical model. Here, nodes denote the constituent random variables variables of the random vectors $X_{t-1}$ and $X_t$. Panel B depicts the unpartitioned system, in which all potential stochastic dependencies of the elements are visualized. The constituent random variables of $X_t$ are conditionally independent given $X_{t-1}$ (cf. \cref{eq:CI}), and the joint distribution $p_{ce}(X_{t-1},X_t)$ is invoked by the assumption of an uncertain marginal distribution $p_u(X_t)$ for each $t = 2,...,T$. Panel C shows an exemplary decomposition of the system, which is based on the bipartition of $(X_{t-1}, X_t)$ into the subsets $\Pi_1^{(i)} = \{a_{t-1},b_{t-1},b_t,c_t\}$ (gray inset) and $\Pi_2^{(i)} = \{a_t,c_{t-1}\}$. In the factorized joint distribution $p_{ce}(X_{t-1},X_t)$, the directed links across the partition boundary are removed, while the links within each partition remain.}\label{fig:Figure_1}
\end{figure}

IIT models the temporal evolution of a system by a discrete time multivariate stochastic process \citep{Cox1977}
\begin{equation}\label{eq:SP}
p(X_1, X_2, ..., X_T).
\end{equation}
In the probabilistic model \eqref{eq:SP}, $X_t, t = 1,...,T$ denotes a finite set of $d$-dimensional random vectors. Here, the limitation to a finite set of discrete time-points is primarily motivated by the eventual goal to apply the concepts of IIT in a data-analytical setting, not by inherent constraints of IIT. Each random vector $X_t$ comprises random variables $x_{t_i}, i = 1,2,...,d \mbox{ } (d \in \mathbb{N})$  that may take on values in one-dimensional outcome spaces $\mathcal{X}_1,\mathcal{X}_2 ..., \mathcal{X}_d$, such that 
\begin{equation}
X_t = (x_{t_1}, x_{t_2}, ..., x_{t_d})^T
\end{equation}
may take on values in the $d$-dimensional outcome space $\mathcal{X} := \prod_{i=1}^d \mathcal{X}_i$. We assume $\mathcal{X} \subseteq \mathbb{R}^d$ throughout. IIT further assumes that the stochastic process fulfils the Markov property, i.e., that the probabilistic model \eqref{eq:SP} factorizes according to
\begin{equation}\label{eq:MP}
p(X_1, X_2, ...,X_T) = p(X_1)\prod_{t = 2}^T p(X_t|X_{t-1}),
\end{equation}
and that the ensuing Markov process is time-invariant, i.e. that all conditional probability distributions $p(X_t|X_{t-1})$ on the right-hand side of \cref{eq:MP} are identical (Figure \ref{fig:Figure_1}A). We will refer to $p(X_t|X_{t-1})$ as the system's \textit{transition probability distribution} in the following. Finally, IIT assumes that the random variables constituting $X_t$ are conditionally independent given $X_{t-1}$, i.e., that the conditional distribution $p(X_t|X_{t-1})$ factorizes according to
\begin{equation}\label{eq:CI}
p(x_{t_1}, x_{t_2}, ...,x_{t_d}|X_{t-1}) = \prod_{i=1}^d p(x_{t_i}|X_{t-1}).
\end{equation}

\subsection{\textcolor{black}{On probabilistic modelling and interventional calculus}}\label{sec:pertub}

\textcolor{black}{Before we lay out our formulation in detail, it is useful to note a subtle but important distinction between the general computational frameworks in \citep{Oizumi2014} and the one used herein. In IIT 3.0, the necessary computations for the definition of integrated information rest on interventional calculus \citep{pearl2009causality}. Specifically, IIT makes use of the $do$ operator, which corresponds to perturbing a system into all possible states and observing the system transitions as a means of assessing the ensuing probability distributions. This has the advantage of being able to define a distribution even if it is not a priori possible to observe every possible system state. However, it has two important disadvantages. First, it leads to the necessity of virtualization to enforce stochastic independence of the system elements, as we will detail below. Second, a perturbational approach poses serious theoretical issues for ultimately transferring the integrated information framework to the case of continuous variables (and specifically Gaussian systems) because there are infinitely many possible states that the system would have to be perturbed into. In contrast, we take the approach of formulating IIT in the realm of general probabilistic models without perturbational calculus and instead express the necessary computations for IIT in terms of a system's joint probability distribution, thereby obviating the need for perturbation and virtualization while facilitating the eventual application of IIT to continuous variables. This corresponds to the fundamental idea that the evaluation of a theoretically derived measure from empirical data can be achieved by estimating the parameters of a probabilistic model from the data, and applying the measure to the thus estimated system model \citep[e.g.,][]{Ostwald2010,Ostwald2014, Ostwald2016}.}

\subsection{Characterization of a system by its joint probability distribution}\label{sec:repertoires}

The stochastic process' forward transition probability distribution is defined as the conditional distribution of $X_t$ given $X_{t-1}$ 
\begin{equation}\label{eq:pe}
p_e(X_t | X_{t-1}) :=  p(X_t|X_{t-1}).
\end{equation}

\noindent Next, we define a joint distribution $p_{ce}$ over $X_{t-1}$ and $X_t$ by multiplication of the Markov transition probability distribution $p(X_t|X_{t-1})$ with a marginal distribution $p_u(X_{t-1})$, i.e.
\begin{equation}\label{eq:pce}
p_{ce}(X_{t-1}, X_{t}) := p_u(X_{t-1})p(X_t|X_{t-1}).
\end{equation}
Here, the marginal distribution $p_u(X_{t-1})$ is meant to represent a maximum of uncertainty about $X_{t-1}$, and for the case of a finite outcome space $\mathcal{X}$, corresponds to the uniform distribution over all states. Note that in \citep{Oizumi2014}, a maximum entropy distribution is defined for the perturbational distribution $p_{per}(X_{t-1})$ used for perturbing the system into all possible states with equal probability. 
Based on the joint distribution of \eqref{eq:pce}, the backward transition probability distribution is then defined as the conditional distribution of $X_{t-1}$ given $X_t$:
\begin{equation}\label{eq:pc}
p_c(X_{t-1} | X_{t}) := \frac{p_{ce}(X_{t-1}, X_{t})}{\sum_{X_{t-1}}p_{ce}(X_{t-1}, X_{t})}
\end{equation}\\

\subsection{Definition of integrated cause-effect information $\phi_{ce}$}

Based on the assumptions of \cref{eq:SP,eq:MP,eq:CI}, IIT defines the integrated cause-effect information $\phi_{ce}$ of a set of system elements in a state $X^* \in \mathcal{X}$ as follows:
\begin{equation}\label{eq:phi}
\phi_{ce} : \mathcal{X} \rightarrow \mathbb{R}, X^* \mapsto \phi_{ce}(X^*) := \min \left\lbrace  \phi_e(X^*),\phi_c(X^*) \right\rbrace,
\end{equation}
where $\phi_e : \mathcal{X} \rightarrow \mathbb{R}$ and $\phi_c : \mathcal{X} \rightarrow \mathbb{R}$ are defined as
\begin{equation}\label{eq:phie}
\phi_e(X^*) := \min_{i \in I} \left\lbrace D \left(p_e(X_{t}|X_{t-1} = X^*) || p_e^{(i)}(X_t|X_{t-1} = X^*) \right)  \right\rbrace
\end{equation}
and
\begin{equation}\label{eq:phic}
\phi_c(X^*) := \min_{i \in I} \left\lbrace D \left(p_c(X_{t-1}|X_t = X^*) || p_c^{(i)}(X_{t-1}|X_t = X^*) \right) \right\rbrace,
\end{equation}
respectively. Note that this applies generally, regardless of whether we consider the whole system $X^d$ or a subset of system elements $X^S \subset X^d$.  
In \eqref{eq:phic} and \eqref{eq:phie}, 
\begin{itemize}
\item $\phi_e(X^*)$ and $\phi_c(X^*)$ are referred to as \textit{integrated effect information} and \textit{integrated cause information} of the state $ X = X^*$,

\item $p_c(X_{t-1}|X_t = X^*)$ and $p_e(X_t|X_{t-1} = X^*)$ are conditional probability distributions that are constructed from the joint probability distribution $p(X_t, X_{t-1})$ of the stochastic process as detailed below and are referred to as the \textit{effect repertoire} and \textit{cause repertoire} of state $X^*$, respectively, 

\item $p_e^{(i)}(X_t|X_{t-1} = X^*)$ and $p_c^{(i)}(X_{t-1}|X_t = X^*)$ are ``decomposed variants'' of the effect and cause repertoires, that result from the removal of potential stochastic dependencies in the system's transition probability distribution as detailed below,

\item $I$ is an index set, the elements of which index the ``decomposed variants'' of the effect and cause repertoires, and 

\item $D : P \times P \rightarrow \mathbb{R}_+, (p_1,p_2) \mapsto D(p_1||p_2)$ denotes a divergence measure between (conditional) probability distributions over the same random entity, with $P$ indicating the set of all possible distributions of this entity. While a variety of distance measures can be used for this assessment in principle (see also \citep{Tegmark2016}), we will in practice follow \citet{Oizumi2014} in defining $D$ as the earth mover's distance for discrete state systems \citep{Levina2001,Mallows1972} due to its increased sensitivity to state differences as compared to the Kullback-Leibler Divergence \citep{Kullback1951}.
\end{itemize}

\noindent We next discuss the intuitive and technical underpinnings of the constituents of the definition $\phi_{ce}$ by \cref{eq:phi,eq:phic,eq:phie} in further detail.

\subsection{System decomposition}\label{sec:SysDecomp}

To evaluate integrated cause-effect information $\phi_{ce}$, IIT first considers all possible ways to decompose a system into two subsets that do not influence each other. The aim is then to identify the system decomposition which, for a given set of system elements in a particular state, is most similar to the actual system in terms of the divergence between the system state's effect and cause repertoires (cf. \cref{eq:phie,eq:phic}). The particular decomposition which fulfills this criterion is labelled the minimum information partition (MIP). In technical terms, the ``system'' to be decomposed corresponds to the collection of random variables and their conditional dependencies that define the discrete time multivariate stochastic process (cf. \cref{eq:SP}). Because of the process' time-invariant Markov property (cf. \cref{eq:MP}), the relevant random variables are the constituents of two time-adjacent random vectors $X_{t-1}$ and $X_t$. As seen above, based on an uncertain marginal distribution over $X_{t-1}$, one may define a joint distribution $p_{ce}(X_{t-1},X_t)$ of these vectors for each $t=2,...,T$. Note that the joint distribution $p_{ce}(X_{t-1},X_t)$ can equivalently be regarded as a joint distribution over the set of all constituent random variables of the random vectors $X_{t-1}$ and $X_t$,
\begin{equation}\label{eq:Xtt}
(X_{t-1},X_t) := \{x_{t-1_1}, x_{t-1_2}, ...,x_{t-1_d},x_{t_1}, x_{1_2}, ...,x_{t_d}\}.
\end{equation}

IIT then uses the intuitive appeal of graphical models \citep{Lauritzen1996, Jordan1998} to introduce the idea of ``cutting a system'' into two independent parts (therefore a \textit{bi}partition). Technically, cutting the graphical model of $p_{ce}(X_{t-1},X_t)$ corresponds to (a) partitioning the set of random variables in \cref{eq:Xtt} into two disjoint subsets and (b) removing all stochastic dependencies across the boundary between the resulting random variable subsets while retaining conditional dependencies within each subset as detailed below (cf. also Figure \ref{fig:Figure_1}B and \ref{fig:Figure_1}C). Notably, there are $k:=2^{2d-1}-1$ unique ways to bipartition a set of cardinality $2d$ (see Appendix for proof). This corresponds to $k$ ways of cutting the corresponding graphical model and thus induces a set of $k$ differently factorized joint distributions $p_{ce}^{(i)}(X_{t-1},X_t), i = 1,...,k$, which form the basis for the decomposed effect and cause repertoires $p_e^{(i)}(X_t|X_{t-1})$ and $p_c^{(i)}(X_{t-1}|X_t)$ in the definition of $\phi_{ce}$ (cf. \cref{eq:phie,eq:phic}).

We next formalize the construction of $p_{ce}^{(i)}(X_{t-1},X_t)$ for $i = 1,...,k$. To this end, first recall that a partition of a set $S$ is a family of sets $P$ with the properties 
\begin{equation}\label{eq:setprop}
\emptyset \notin P, \bigcup_{M\in P} M = S, \mbox{ and if } M,M' \in P \mbox{ and } M \neq M', \mbox{ then } M \cap M' = \emptyset.
\end{equation}
Let $\Pi^{(i)}$ denote a bipartition of a subset of random variables $(X^{S}_{t-1},X^{S}_t)$ under scrutiny, i.e.
\begin{equation}\label{eq:BP1}
\Pi^{(i)} := \left(\Pi_1^{(i)},\Pi_2^{(i)} \right),
\end{equation}
where
\begin{equation}\label{eq:BP2}
\Pi_1^{(i)},\Pi_2^{(i)} \subset (X^{S}_{t-1},X^{S}_t), \Pi_1^{(i)} \cap \Pi_2^{(i)} = \emptyset \mbox{ and } \Pi_1^{(i)} \cup \Pi_2^{(i)} = (X^{S}_{t-1},X^{S}_t). 
\end{equation}
Let further
\begin{equation}\label{eq:pBP1BP2}
p_{ce}(\Pi_1^{(i)}) = \sum_{\Pi_2^{(i)}} p_{ce}(X^{S}_{t-1},X^{S}_t) \mbox{ and }
p_{ce}(\Pi_2^{(i)}) = \sum_{\Pi_1^{(i)}} p_{ce}(X^{S}_{t-1},X^{S}_t) 
\end{equation}
denote the marginal distributions of $p_{ce}(X^{S}_{t-1},X^{S}_t)$ (cf \cref{eq:pce}) of the random variables contained in $\Pi_1^{(i)}$ and $\Pi_1^{(i)}$, respectively. Then the elements of the set of factorized variants of the joint distribution $p_{ce}(X^{S}_{t-1},X^{S}_t)$ are given by 
\begin{equation}\label{eq:pcefact}
p_{ce}^{(i)}\left(X^{S}_{t-1},X^{S}_t\right) := p_{ce}(\Pi_1^{(i)})p_{ce}(\Pi_2^{(i)}) \mbox{ for } i = 1,2,..,k.
\end{equation}

\color{black} When partitioning a system into two independent parts, the perturbational approach in IIT necessitates the introduction of virtual elements in the calculation of the ensuing conditional probability distributions. Therefore, we now aim to clarify what this means in the context of a formulation based on the system's joint probability distribution and then provide a general formula for the evaluation of these repertoires in both the partitioned and the unpartitioned case.

\subsection{Virtualization is factorization}\label{sec:comp_exclu}

\begin{figure}[ht!]
\centering
\includegraphics[width = 85 mm]{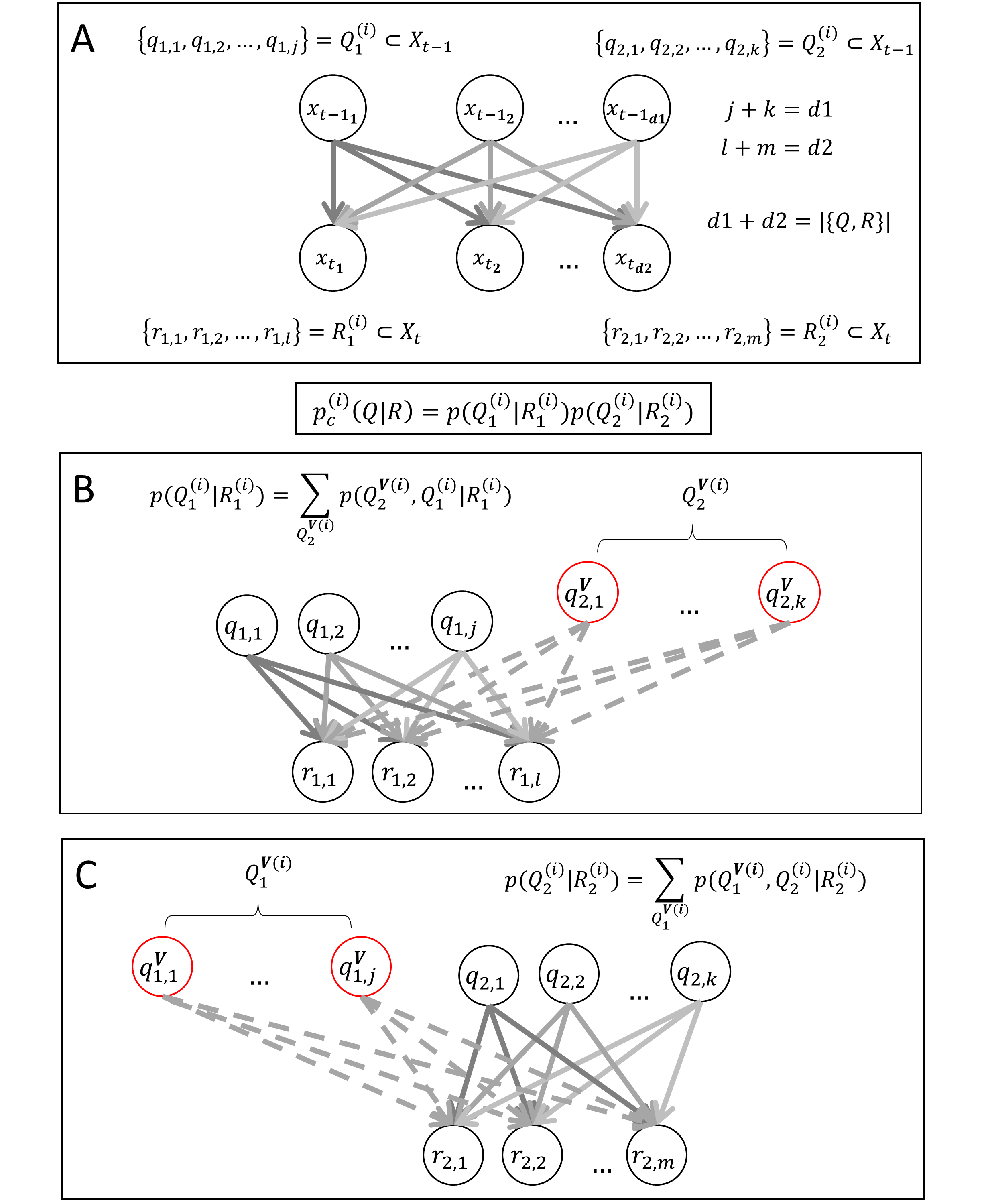}
 \caption{\textbf{General system decomposition and virtualization in IIT 3.0 \cite{Oizumi2014}}. \textcolor{black}{Panel A visualizes the system decomposition in a general manner for the cause repertoire. The cause repertoire is decomposed as a factorization of two conditional distributions for every partition $i$ (see equation inset below panel A). For unique reference to \citep{Oizumi2014} (Supplements, text S2), we denote the elements in the partition of a subset of $X_{t-1}$ with cardinality $d1$ by $\{q_{1,1},...,q_{1,j},q_{2,1},...,q_{2,k}\}=Q$, where $j+k=d1$. Likewise, the elements in the partition of a subset of $X_t$ with cardinality $d2$ are denoted by $\{r_{1,1},...,r_{1,l},r_{2,1},...,r_{2,m}\}=R$, where $l+m=d2$. Panel B shows the first conditional distribution, and Panel C the second conditional distribution of the factorization. Virtualization is indicated by the superscript $V$. Every element in $Q^{V(i)}_2$ comprises $l$ virtual elements with independent connections to each element in $R^{i}_1$, and likewise for $Q^{V(i)}_1$, yielding a total of $l*k$ virtual elements in the former and $j*m$ in the latter virtual set. Every red circle in panels B and C thus summarizes a set of independent virtual elements connected to the respective elements in the partition of $R$. As we show in the main text, however, the necessity of virtualization stems from the perturbational approach of \citep{Oizumi2014}, and the resulting distributions are equivalently found by factorization of the system's joint distribution.}}\label{fig:Figure_virt}
\end{figure}

The intuition behind virtualization is to account for correlation effects over the subset of variables in question due to common input from outside of this considered subset. In order to decorrelate this common input, virtual elements are introduced with independent output to the elements inside of the considered subsystem, and a maximum entropy distribution is defined over the input states of these virtual elements. Note that this is a consequence of the perturbational calculations. If, for instance, a system element $x_{t-1_1}$ provides input to two elements $x_{t_1}$ and $x_{t_2}$, then perturbing the state of $x_{t-1_1}$ will indeed lead to correlations between $x_{t_1}$ and $x_{t_2}$ because the input (i.e. the state of $x_{t-1_1}$) will automatically affect both $x_{t_1}$ and $x_{t_2}$ due to the connectivity of the system. If we are to assess the effect that the state of $x_{t-1_1}$ has on $x_{t_1}$ and $x_{t_2}$ independently, however, we must remove this correlation. One solution is to define two "virtual elements" $x^{V1}_{t-1_1}$ and $x^{V2}_{t-1_1}$ that can be perturbed independently, thereby effectively removing the stochastic dependence of these variables (or "noising the connections" between them). Formally, the idea behind virtualization is thus to enforce conditional independence on the variables within a subset in question from elements outside this subset. In the following, however, we ascertain that there is no need to introduce the concept of virtualization because 1) "inputs" from one element to another have an implicit temporal direction (input always refers to the previous temporal state), 2) virtual elements and real past elements share the same state space and a maximum entropy marginal distribution is placed over virtual elements just as over past states (cf. \cref{eq:pce}), and 3) in calculating the actual probability distributions, we always marginalize over virtual elements, thus leading to the same output distributions. Instead, we aim to show that virtualization corresponds to the factorization of the system's joint distribution as defined in \cref{eq:pce}.  

\color{black}
\noindent Figure \ref{fig:Figure_virt} shows the system decomposition in IIT 3.0 for the cause repertoire along with the virtualization, which we will denote in the following by the superscript $V$. For explicit reference (cf. supplementary text S2 in \citep{Oizumi2014}), we refer to the numerator by $Q^{i}$ (the inputs) and to the denominator by $R^{i}$ which are partitioned to $Q_1^{(i)}$, $Q_2^{(i)}$ and $R_1^{(i)}$, $R_2^{(i)}$, respectively, depending on partition $i$.  
The cause repertoire is factorized according to

\begin{equation}\label{eq:Virt1}
p_{c}^{(i)}(Q|R) = p(Q_1^{(i)}|R_1^{(i)})p(Q_2^{(i)}|R_2^{(i)})
\end{equation}
For a system subset $X = X^S$ under consideration ("purview") with $|X^S_{t-1}| = d1$, $|X^S_{t}| = d2$ and $d1 + d2 = |\{Q,R\}|$ (see fig. \ref{fig:Figure_virt}),
\begin{equation}\label{eq:Virt2}
Q_1^{(i)},Q_2^{(i)}\subset{X^S_{t-1}}, Q_1^{(i)} \cap Q_2^{(i)} = \emptyset \mbox{ and } Q_1^{(i)} \cup Q_2^{(i)} = Q^{(i)} = X^S_{t-1}. 
\end{equation}
Similarly,
\begin{equation}\label{eq:Virt3}
R_1^{(i)},R_2^{(i)}\subset{X^{S}_t}, R_1^{(i)} \cap R_2^{(i)} = \emptyset \mbox{ and } R_1^{(i)} \cup R_2^{(i)} = R^{(i)} = X^{S}_t. 
\end{equation}
\noindent For each of the two subsets, virtual elements are introduced over the complement of the respective partition of $Q^{(i)}$ with regard to $X^{S}_{t-1}$ (i.e. over the "inputs" outside of the subset in question), i. e.   
\begin{equation}\label{eq:Virt4}
Q_1^{V(i)} = X^{S}_{t-1}\setminus{Q_1^{(i)}} \mbox{ and } Q_2^{V(i)} = X^{S}_{t-1}\setminus{Q_2^{(i)}}. 
\end{equation}

\noindent Note, however, that due to the perturbational approach, for every element $q^{V(i)}_{1,1},...,q^{V(i)}_{1,l}$ in $Q^{V(i)}_1$, there are in fact $m$ individual virtual elements because perturbation requires a single independent input element from $Q^{V(i)}_1$ to $R^{(i)}_2$, and in analogy for the connections from $Q^{V(i)}_2$ to $R^{(i)}_1$. In \cref{fig:Figure_virt}, we summarize this as a single red circle for every set of independent virtual elements for visual coherence. For every input element in $Q^{V(i)}_1$, IIT places a maximally uncertain perturbational distribution over its states, and likewise for $Q^{V(i)}_2$ (cf. maximum entropy distribution over past states $p_u(X_{t-1})$ in \cref{eq:pce}). \noindent We now form the joint distribution for the two factorized conditional distributions in \cref{eq:Virt1} (cf. \cref{fig:Figure_virt}, panels B and C) as
\begin{equation}\label{eq:Virt6}
\begin{split}
p(Q_1^{(i)},R_1^{(i)}) = & \sum_{Q_1^{V(i)}}p(Q_1^{V(i)},Q_1^{(i)}|R_1^{(i)})p(R_1^{(i)}) \\ = & \sum_{Q_1^{V(i)}}p(Q_1^{V(i)},Q_1^{(i)},R_1^{(i)})
\end{split}
\end{equation}


\noindent and equivalently for $p(Q_2^{(i)},R_2^{(i)})$. Note that we sum over all virtual elements to obtain this subjoint distribution, and that with \cref{eq:setprop,eq:BP1,eq:BP2,eq:pBP1BP2,eq:pcefact} we have 
\begin{equation}\label{eq:Virt7}
\Pi_1^{(i)} = Q_1^{(i)}\cup{R_1^{(i)}} \mbox{ and } \Pi_2^{(i)} =Q_2^{(i)}\cup{R_2^{(i)}}. 
\end{equation}
\noindent With the above and by forming the joint distribution in \cref{eq:Virt1}, we state that  
\begin{align}\label{eq:Virt8}
\begin{split}
p_{ce}^{(i)}(X^{S}_{t-1},X^{S}_t)&=p_{ce}^{(i)}(Q,R)\\
&=p(Q_1^{(i)},Q_2^{(i)},R_1^{(i)},R_2^{(i)})\\ &= \sum_{Q_1^{V(i)}}p(Q_1^{V(i)},Q_1^{(i)},R_1^{(i)})\sum_{Q_2^{V(i)}}p(Q_2^{V(i)},Q_2^{(i)},R_2^{(i)})\\
&=p(Q_1^{(i)},R_1^{(i)})p(Q_2^{(i)},R_2^{(i)})\\
&=p(\Pi_1^{(i)})p(\Pi_2^{(i)})
\end{split}
\end{align}
\noindent where the last equation is the expression stated \cref{eq:pcefact}. The equality for the effect repertoire follows in analogy, with the difference that we now condition $p(Q^{(i)},R^{(i)})$ on $p(Q^{(i)})$, with which \cref{eq:Virt6} becomes
\begin{align}\label{eq:Virt9}
\begin{split}
p(Q_1^{(i)},R_1^{(i)}) &= \sum_{Q_1^{V(i)}}p(R_1^{(i)}|Q_1^{(i)},Q_1^{V(i)})p(Q_1^{(i)},Q_1^{V(i)}) \\ &=\sum_{Q_1^{V(i)}}p(Q_1^{V(i)},Q_1^{(i)},R_1^{(i)})
\end{split}
\end{align}

\noindent and equivalently for $p(Q_2^{(i)},R_2^{(i)})$. 
Note that the above subjoint distribution is identical to \cref{eq:Virt6}, and thus the equivalence in \cref{{eq:Virt8}} follows in analogy.

\subsection{Factorization and distribution normalization}\label{sec:comp_exclu}

Apart from partitioning, the application of virtualization in IIT also concerns the calculation of cause and effect repertoires over a subset $X^S_t \subset X^{d}_t$, where the maximum cardinality of $S$ is $d$ (the whole system of interest). Similarly, $X^S_{t-1} \subset X_{t-1}$ (but note that we do not necessarily refer to the same variables in the subset $X^S_{t-1}$ and $X^S_{t}$). The ensuing subjoint distribution $p(X^S_{t},X^S_{t-1})$ is found from the original joint distribution by marginalizing over the complement of the subset with regard to the whole system, i.e. $X^d_{t}\setminus X^S_t$ and $X^d_{t-1}\setminus X^S_{t-1}$. The aim of virtualization is again to enforce the independence of system elements at time $t$ given their respective inputs. For the case of the effect repertoire, this corresponds to the independence of  $x^S_{t_1}, x^S_{t_2}, ...,x^S_{t_S}$ given $X^S_{t-1}$. For every element in $X^S_{t}$, virtual elements are introduced over the complement of $X^S_{t-1}$ with regard to $X$. Similar to the above, however, the necessary independence is equally enforced by marginalization and multiplication of the ensuing subjoint distributions 
\begin{align}\label{eq:eff_fact}
\begin{split}
p_e(X^S_{t}|X^S_{t-1}) &= \prod_{i=1}^{|S|} \frac{\sum_{X^S_{t}\setminus x^S_{t_i}}p(X^S_{t},X^S_{t-1})}{\sum_{X^S_{t}}
p(X^S_{t},X^S_{t-1})}\\
&= \prod_{i=1}^{|S|} \frac{p(x^S_{t_i},X^S_{t-1})}{
p(X^S_{t-1})}\\
&= \prod_{i=1}^{|S|} p(x^S_{t_i}|X^S_{t-1})
\end{split}
\end{align}
The above essentially corresponds to the assumption of conditional independence inherent to the system model in \cref{{eq:CI}}. Note also that in the absence of any constraint from the past system state, the expression in \cref{eq:eff_fact} reduces to 
\begin{align}\label{eq:max_entropy_forward}
\begin{split}
p_u(X^S_{t})&= \prod_{i=1}^{|S|} p(x^S_{t_i})
\end{split}
\end{align}
which represents the definition of a maximum entropy distribution in the forward temporal direction ("unconstrained future repertoire" in IIT).

For the cause repertoire, we again enforce independence of the elements in $X^S_{t}$ based on their respective inputs in $X^S_{t-1}$. However, we now condition the subjoint $p(X^S_{t},X^S_{t-1})$ on $X^S_{t}$ (intuitively, enforcing "backward" conditional independence), which again corresponds to the factorization of the joint distribution into the corresponding subjoint distributions and forming their product
\begin{align}\label{eq:cause_fact}
\begin{split}
p_e(X^S_{t-1}|X^S_{t}) &= \prod_{i=1}^{|S|} \frac{\sum_{X^S_{t}\setminus x^S_{t_i}}p(X^S_{t},X^S_{t-1})}{\sum_{X^S_{t}\setminus x^S_{t_i}}\sum_{X^S_{t-1}}
p(X^S_{t},X^S_{t-1})}\\
&= \prod_{i=1}^{|S|} \frac{p(x^S_{t_i},X^S_{t-1})}{
p(x^S_{t_i})}\\
&= \prod_{i=1}^{|S|} p(X^S_{t-1}|x^S_{t_i})
\end{split}
\end{align}
Based on \cref{eq:eff_fact} and \cref{eq:cause_fact}, there are a couple of interesting aspects to mention. First, note that in the second line of both equations, the subjoint distribution in the numerator is the same and all necessary distributions are easily obtained from the whole system's joint distribution. Second, we can state a general rule of when repertoire normalization is necessary in IIT. This will be the case for the cause repertoire if
\begin{equation}\label{eq:rule_norm}
\begin{split}
\prod_{i=1}^{|S|} p(x^S_{t_i}) \neq \sum_{X^S_{t-1}}p(X^S_{t},X^S_{t-1})\\
\end{split}
\end{equation}
i.e. depending on whether it makes a difference to the cause repertoire if the marginal over $X^S_{t}$ factorizes or not. If it does, the cause repertoire must be normalized by the sum over all previous states $X^{S}_{t-1}$ for every current state to ensure unity, i.e. $\sum_{X^{S}_{t-1}}\prod_{i=1}^{|S|} p(X^S_{t-1}|x^S_{t_i} = x^{S*}_{t_i})$, which, computationally, corresponds to column-wise matrix normalization and is equivalent to the formulations in \citep{Tononi2015,marshall2016integrated}. Note that the effect repertoire in \cref{eq:eff_fact} is always conditioned on the marginal $p(X^S_{t-1})$, and thus never needs to be normalized. Third, if the cardinality of $X^S_{t}$ is 1, i.e. we assess the cause repertoire over a single variable $x^S_{t}$, then the inequality in \cref{eq:rule_norm} is never true, which means that these repertoires never require normalization and which is also the reason why virtualization (i.e. factorization) is necessary for "higher order mechanisms" in IIT (see cause repertoire in text S2, \citep{Oizumi2014}).   
Finally, note that based on the system decomposition related above, we factorize the system's joint distribution into two subjoint distributions $p_{ce}(\Pi_1^{(i)})$ and $p_{ce}(\Pi_2^{(i)})$ in order to induce independence between the corresponding two subsets of variables. In evaluating the cause and effect repertoires of the partitioned system, we then factorize $p_{ce}(\Pi_1^{(i)},\Pi_2^{(i)})$(cf. \cref{eq:pcefact}) again, according to \cref{eq:eff_fact} and \cref{eq:cause_fact}. To this end, let \begin{equation}\label{eq:Pi_t}
\Pi^{(i)}_{1,t} = \Pi^{(i)}_1 \cap X^S_t \mbox{, } \Pi^{(i)}_{2,t} = \Pi^{(i)}_2 \cap X^S_t \mbox{, and naturally } \Pi^{(i)}_{t} = \Pi^{(i)}_{1,t} \cup \Pi^{(i)}_{2,t} = X^S_{t}.
\end{equation}
and equally   
\begin{equation}\label{eq:Pi_t-1}
\Pi^{(i)}_{1,t-1} = \Pi^{(i)}_1 \setminus \Pi^{(i)}_{1,t} \mbox{, } \Pi^{(i)}_{2,t-1} = \Pi^{(i)}_2 \setminus \Pi^{(i)}_{2,t} \mbox{ and } \Pi^{(i)}_{t} = \Pi^{(i)}_{1,t-1} \cup \Pi^{(i)}_{2,t-1} = X^S_{t-1}.
\end{equation}
Let $z_1$ and $z_2$ denote the cardinality of $\Pi^{(i)}_{1,t}$ and $\Pi^{(i)}_{2,t}$, respectively. Note that if $z_1 = 0$, then $z_2 = |X^S_t|$, and vice versa, and always $z_1 \cup z_2 = z = |X^{S}_t|$. Similarly, let $u_1 = |\Pi^{(i)}_{1,t-1}|$ and $u_2 = |\Pi^{(i)}_{2,t-2}|$. We now apply the general formulas in \cref{eq:eff_fact,eq:cause_fact} to $p_{ce}(\Pi_1^{(i)},\Pi_2^{(i)})$ by defining 
\begin{equation}\label{eq:fact_part_joint}
p_{ce}(\Pi^{(i)}_{t},\Pi^{(i)}_{t-1}) :=   
\begin{cases} 
\prod_{h=1}^{z_2} p(\Pi^{(i)}_{2,t,h},)p(\Pi^{(i)}_{1,t-1}) & \mbox{, if } z_1 = 0, u_2 = 0 \\
\prod_{h=1}^{z_2} p(\Pi^{(i)}_{2,t,h},\Pi^{(i)}_{2,t-1})p(\Pi^{(i)}_{1,t-1}) & \mbox{, if } z_1 = 0, u_{1,2} \neq 0 \\
\prod_{h=1}^{z_1} p(\Pi^{(i)}_{1,t,h})p(\Pi^{(i)}_{2,t-1}) & \mbox{, if } z_2 = 0, u_{1} = 0 \\
\prod_{h=1}^{z_1} p(\Pi^{(i)}_{1,t,h},\Pi^{(i)}_{1,t-1})p(\Pi^{(i)}_{2,t-1}) & \mbox{, if } z_2 = 0, u_{1,2} \neq 0 \\
\prod_{h=1}^{z_1} p(\Pi^{(i)}_{1,t,h})\prod_{h=1}^{z_2} p(\Pi^{(i)}_{2,t,h},\Pi^{(i)}_{2,t-1}) & \mbox{, if } z_{1,2} \neq 0, u_1 = 0\\ 
\prod_{h=1}^{z_1} p(\Pi^{(i)}_{1,t,h},\Pi^{(i)}_{1,t-1})\prod_{h=1}^{z_2} p(\Pi^{(i)}_{2,t,h}) & \mbox{, if } z_{1,2} \neq 0, u_2 = 0\\
\prod_{h=1}^{z_1} p(\Pi^{(i)}_{1,t,h},\Pi^{(i)}_{1,t-1})\prod_{h=1}^{z_2} p(\Pi^{(i)}_{2,t,h},\Pi^{(i)}_{2,t-1}) & \mbox{, if } z_{1,2} \neq 0, u_{1,2} \neq 0. 
\end{cases} 
\end{equation}
Intuitively, we thus factorize the subjoints $p(\Pi^{(i)}_1)$ and $p(\Pi^{(i)}_2)$ into as many factors as they contain variables in $X^S_t$, where the case distinction above accounts for the marginal cases in which one of the subsets is empty due to the partition $i$. Consider for example the case of a two-dimensional random vector $X_t$ and a bipartition of the form $\Pi_1 = \{x_{{t-1}_1}\}$ and $\Pi_2 = \{x_{{t-1}_2},x_{t_1}, x_{t_2}\}$. Taking the approach in \citep{Oizumi2014} leads to empty conditionals, here: $p([    ]|x_{{t-1}_1})p(x_{t_1}, x_{t_2}|x_{{t-1}_2})$ 
for the partitioned effect repertoire (which requires the assumption that $p(x|[    ]) = p(x)$ and $p([    ]|x) = 1$ to recover well-definedness). The above distinction eschews this. For instance, the example partition corresponds to the factorization $p(x_{t-1_1})p(x_{t_1},x_{t-1_2})p(x_{t_2},x_{t-1_2})$ because $z_1 = 0, u_{1,2} \neq 0$.\\ 
With \cref{eq:fact_part_joint}, we now have a general rule to factorize the system's joint distribution. In order to state a general rule to calculate the cause and effect repertoires, however, we still need to condition the thus factorized joint distribution on the corresponding marginal distribution. For the effect repertoire, this marginal is given by
\begin{equation}\label{eq:marg_effect_fact}
p_{ce}(\Pi^{(i)}_{t-1}) :=  
\begin{cases} 
\prod^{z_1}p(\Pi^{(i)}_{1,t-1}) & \mbox{, if } u_2 = 0 \\
\prod^{z_2}p(\Pi^{(i)}_{2,t-1}) & \mbox{, if } u_1 = 0 \\
\prod^{z_1}p(\Pi^{(i)}_{1,t-1})\prod^{z_2}p(\Pi^{(i)}_{2,t-1}) & \mbox{, if } u_{1,2} \neq 0. 
\end{cases} 
\end{equation} For the cause repertoire, the marginal always factorizes (cf. \cref{eq:cause_fact}) and is thus
\begin{equation}\label{eq:marg_cause_fact}
p_{ce}(\Pi^{(i)}_{t}) := \prod_{h=1}^{z_1} p(\Pi^{(i)}_{1,t,h})\prod_{h=1}^{z_2} p(\Pi^{(i)}_{2,t,h}) = \prod_{i=1}^{|S|} p(x^S_{t_i})
\end{equation}
We thus generally state that for every partition $i$, the cause repertoire is given by
\begin{equation}\label{eq:bpcr}
p_c^{(i)}(X^S_{t-1}|X^S_t) = \frac{p_{ce}(\Pi^{(i)}_{t},\Pi^{(i)}_{t-1})}{p_{ce}(\Pi^{(i)}_{t})}
\end{equation}
and the effect repertoire by
\begin{equation}\label{eq:bpce}
p_e^{(i)}(X^S_{t}|X^S_{t-1}) = \frac{p_{ce}(\Pi^{(i)}_{t},\Pi^{(i)}_{t-1})}{p_{ce}(\Pi^{(i)}_{t-1})}.
\end{equation}

\color{black}

\noindent \textit{Example}

\noindent For a brief illustration of the above, we consider the exemplary system of Figure \ref{fig:Figure_1}. Here, the concatenated state vector over two adjacent time-points is given by (cf. \cref{eq:Xtt})
\begin{equation}\label{eq:Xttex}
(X_{t-1},X_t) = \{a_{t-1},b_{t-1},c_{t-1},a_t, b_t, c_t\}
\end{equation}
One of the $k = 2^{6-1} - 1 = 31 $ bipartitions of \cref{eq:Xttex} (which we label here as $i:=1$) is given by
\begin{equation}\label{eq:Piex}
\Pi_1^{(1)} = \{a_{t-1},b_{t-1},b_t,c_t\} \mbox{ and } 
\Pi_2^{(1)} = \{c_{t-1},a_t\}.  
\end{equation}
Note that this corresponds to the partition depicted in panel B of Figure \ref{fig:Figure_1}. Hence, with \cref{eq:pBP1BP2}
\begin{equation}
p_{ce}(\Pi_1^{(1)}) = p_{ce}(a_{t-1},b_{t-1},b_t,c_t) \mbox{ and } 
p_{ce}(\Pi_2^{(1)}) = p_{ce}(c_{t-1},a_t).
\end{equation}
We have $\Pi^{(1)}_{1,t-1} = \{a_{t-1},b_{t-1}\}$, $\Pi^{(1)}_{1,t} = \{b_{t},c_{t}\}$, $\Pi^{(1)}_{2,t-1} = \{c_{t-1}\}$, and $\Pi^{(1)}_{2,t} = \{a_{t}\}$. Thus, $z_{1,2} \neq 0, u_{1,2} \neq 0$, which yields (\cref{eq:fact_part_joint}) the fully factorized joint distribution  
\begin{equation}
p_{ce}(\Pi^{(1)}_{t},\Pi^{(1)}_{t-1}) = p_{ce}(a_{t-1},b_{t-1},b_t)p_{ce}(a_{t-1},b_{t-1},c_t)p_{ce}(c_{t-1},a_t)
\end{equation} and based on \cref{eq:marg_effect_fact,eq:marg_cause_fact} the marginal distributions
\begin{align}
\begin{split}
& p_{ce}(\Pi^{(1)}_{t-1}) = p_{ce}(a_{t-1},b_{t-1})p_{ce}(a_{t-1},b_{t-1})p_{ce}(c_{t-1}) \mbox {, and }\\ & p_{ce}(\Pi^{(1)}_{t}) = p_{ce}(a_{t})p_{ce}(b_{t})p_{ce}(c_{t}). 
\end{split}
\end{align}
The decomposed cause repertoire is then given by \cref{eq:bpcr} as 
\begin{align}
\begin{split}
\frac{p_{ce}(\Pi^{(1)}_{t},\Pi^{(1)}_{t-1})}{p_{ce}(\Pi^{(1)}_{t})} & = \frac{p_{ce}(a_{t-1},b_{t-1},b_t)p_{ce}(a_{t-1},b_{t-1},c_t)p_{ce}(c_{t-1},a_t)}{p_{ce}(a_{t})p_{ce}(b_{t})p_{ce}(c_{t})}\\
& = p_{ce}(a_{t-1},b_{t-1}|b_t)p_{ce}(a_{t-1},b_{t-1}|c_t)p_{ce}(c_{t-1}|a_t) 
\end{split}
\end{align} requiring normalization by the sum over $p(\Pi_{t-1}|\Pi_{t}=\Pi^*_{t})$, and the decomposed effect repertoire (\cref{eq:bpce}) evaluates to
\begin{align}
\begin{split}
\frac{p_{ce}(\Pi^{(1)}_{t},\Pi^{(1)}_{t-1})}{p_{ce}(\Pi^{(1)}_{t-1})} & = \frac{p_{ce}(a_{t-1},b_{t-1},b_t)p_{ce}(a_{t-1},b_{t-1},c_t)p_{ce}(c_{t-1},a_t)}{p_{ce}(a_{t-1}, b_{t-1})p_{ce}(a_{t-1}, b_{t-1})p_{ce}(c_{t-1})}\\
& = p_{ce}(b_t|a_{t-1},b_{t-1})p_{ce}(c_t|a_{t-1},b_{t-1})p_{ce}(a_t|c_{t-1}) 
\end{split}
\end{align}

\noindent For further illustration of this constructive process, an exhaustive example is provided in the supplements, where the effect and cause repertoires corresponding to the seven bipartitions of a two-dimensional system are considered in detail.

\color{black}
\subsection{On composition and exclusion}\label{sec:comp_exclu}

One of the main theoretical advances of IIT 3.0 over previous formulations is the extension of the general framework to exclude superposition of multiple causes and effects (exclusion principle) and to reflect the composition of the system in the definition of integrated information on a system level (composition principle).
To this end, the evaluation of $\phi_c$ and $\phi_e$ as specified above is carried out in two distinct ways over the powerset of the system elements.

\subsubsection{Exclusion principle}\label{sec:exclusion}

The intuition behind the exclusion principle is that just as any conscious experience excludes all others, 
in physical systems sustaining consciousness, causes and effects must not be "multiplied beyond necessity" and only maximally integrated cause and effect repertoires of a set of elements can contribute to consciousness, thereby excluding all other possible causes and effects \citep{Oizumi2014}.  
Mathematically, for a given subset $ X^S \subset X^d$, the evaluation of $\phi_c$ and $\phi_e$ is therefore carried out over all possible cause and effect repertoires, which are specified by the powerset of the system elements. Excluding the empty set, the system's powerset is generally given by 
\begin{equation}\label{eq:powerset_X}
\mathcal{P}(X^d) = \lbrace \lbrace x_{1}\rbrace , \lbrace x_{2} \rbrace, ..., \lbrace x_{d} \rbrace, \lbrace x_{1}, x_{2} \rbrace, ...,  \lbrace x_{1}, x_{2}, ...,  x_{d} \rbrace  \rbrace 
\end{equation}
\noindent with cardinality $C = 2^d -1$. 
For notational clarity, let every subset in the powerset be denoted by $ \mathcal{P}(X) := \lbrace \lbrace X^{\mathcal{P}_1}\rbrace , \lbrace X^{\mathcal{P}_2} \rbrace, ..., \lbrace X^{\mathcal{P}_C} \rbrace  \rbrace$.
For a given subset $X^S_t \subset X^d_t$, we thus compute a total of $C$ cause and $C$ effect repertoires.
The set of cause repertoires for $X^S_t$ is thus given by 
\begin{equation}\label{eq:powerset_cause}
p_{c}^{(j)}\left(\mathcal{P}(X_{t-1})|X^S_t\right) := p_{c}(X^{\mathcal{P}_j}_{t-1}| X^S_t)
\end{equation}
and the set of effect repertoires by 
\begin{equation}\label{eq:powerset_effect}
p_{e}^{(j)}\left(\mathcal{P}(X_{t+1})|X^S_t\right) := p_{e}(X^{\mathcal{P}_j}_{t+1}| X^S_t)
\end{equation}
with $j = 1,2,...,C$.
\noindent For illustration, consider the thus defined set of cause repertoires for the case $X^S_t = x_{t_1}$. We thus compute $p_{c}(X^{\mathcal{P}_j}_{t-1}| x_{t_1})$, or, explicitly, the distributions $p(x_{t-1_1}| x_{t_1}), p(x_{t-1_2}| x_{t_1}), ..., p(x_{t-1_1}, x_{t-1_2}| x_{t_1}), ..., p(X_{t-1}| x_{t_1})$.\\ 
Through system decomposition, we obtain a total of $C$ different $\phi_c$ and $\phi_e$ values, one for every decomposition of the $j$'th cause and effect repertoires. Of all those $\phi_c$ and $\phi_e$ values obtained over the powerset, the exclusion postulate in IIT 3.0 now requires that only the maximally integrated cause (and respectively effect) information be considered. 
\begin{equation}\label{eq:max_small_phi}
\phi_{c}^{max} := \max_{j \in C} \left\lbrace \phi_{c}^{j} \right\rbrace \mbox{, and } \phi_{e}^{max} := \max_{j \in C} \left\lbrace \phi_{e}^{j} \right\rbrace
\end{equation}
The cause repertoire $p_{c}(X^{\mathcal{P}_j=j^*}_{t-1}| X^S_t)$ whose decomposition yields $\phi_{c}^{max}$ is called the maximally integrated cause repertoire of $X^S_t$ (recall that this is always evaluated for $X^S_t$ being in a particular state), and equivalently for the maximally integrated effect repertoire. Here, $j^*$ refers to the corresponding subset of the powerset (note that $j^*$ does not have to be the same for $\phi_{c}^{max}$ and $\phi_{e}^{max}$). 
The minimum of maximally integrated cause and maximally integrated effect information then defines maximally integrated cause-effect information \begin{equation}\label{eq:max_cei}
\phi_{ce}^{max} := \min \left\lbrace \phi_{c}^{max},\phi_{e}^{max} \right\rbrace
\end{equation}
If a subset of $X^S_t$ being in a particular state specifies $\phi_{ce}^{max}>0$, it forms a maximally irreducible cause-effect repertoire (a "concept" in \citep{Oizumi2014}). Notably, in the IIT framework, this concept is identical to a quale in the strict sense of the word. Intriguingly, the particular repertoire $j^*$ yielding $\phi_{c}^{max}$ (and equivalently for $\phi_{e}^{max}$) is not necessarily unique. While this may seem like a mathematical detail at this point, it has important implications both for the quantification of capital $\Phi$ (see below) and the interpretation of a concept as a point in qualia space (see discussion). 

\subsubsection{Composition principle}\label{sec:composition}

The composition principle is a natural extension of the above. By iterating over all possible cause and effect repertoires for a subset of $X^S_t$ being in a particular state, we define $\phi_{ce}^{max}$ for that subset in its state. In order to take system composition into account, we now compute $\phi_{ce}^{max}$ not only for a specific subset $X^S_t$ but rather over all possible subsets of the system $X^d_t$, i.e. again over the powerset. For every element $j$ in the powerset, we thus compute the set of cause repertoires as
\begin{equation}\label{eq:powerset_composition_cause}
p_{c}^{(j)}\left(\mathcal{P}(X_{t-1})|X^{\mathcal{P}_j}\right) := p_{c}(\mathcal{P}(X_{t-1})| X^S_{t} = X^{\mathcal{P}_j}_t)
\end{equation}
and the set of effect repertoires as 
\begin{equation}\label{eq:powerset_composition_effect}
p_{e}^{(j)}\left(\mathcal{P}(X_{t+1})|X^{\mathcal{P}_j}\right) := p_{e}(\mathcal{P}(X_{t+1})| X^S_{t} = X^{\mathcal{P}_j}_t)
\end{equation}
for $j = 1,2,..,C$. We thus obtain a total of $C$ values for $\phi_{ce}^{max}$. Together, all those subsets $X^{\mathcal{P}_j}$ that specify a maximally integrated cause-effect repertoire are considered a "conceptual structure" in IIT, i.e. a set of concepts. In the following, let the number of concepts be denoted by $J^*$.

\subsection{Integrated conceptual information $\Phi$}\label{sec:composition}

We are now in a position to define the integrated information capital $\Phi$ of the conceptual structure of a system $X^d$ being in a particular state $X^{d*}$. The idea behind $\Phi$ is to quantify how much a constellation of concepts specified by a system state is irreducible to its individual parts. Formally, this corresponds to quantifying how much the information inherent in a system's state conceptual structure can be reduced. Thus, we first need to define the conceptual information $CI$ that is specified by the constellation of concepts. IIT defines this as the sum of the distances between a maximally integrated cause and effect repertoire to the respective maximum entropy distribution in the past or future (cf. \cref{eq:pce} and \cref{eq:max_entropy_forward}), weighted by their $\phi_{ce}^{max}$ values, for all $J^*$ concepts that a system $X^d$ in state $X^{d*}$ specifies:  
\begin{equation}\label{eq:concept_info}
\begin{split}
CI(X^{d*}_t) := \sum^{J^*}_{j^*=1}
\phi_{ce}^{max,j^*}\left(D\left(p_{c}^{(j^*)}(X^{\mathcal{P}_j^{*}}_t)||p_u(X_{t-1})\right) +D\left(p_{e}^{(j^*)}(X^{\mathcal{P}_j^{*}}_t)||p_u(X_{t+1})\right)\right)
\end{split}
\end{equation}
However, due to the aspect of non-unique maximally integrated cause and effect repertoires (which we will illustrate in a discrete state example system below), we instead define the conceptual information of a constellation of concepts simply as the sum of all $\phi_{ce}^{max}$ values of those concepts 
\begin{equation}\label{eq:concept_info_us}
\begin{split}
CI(X^{d*}_t) := \sum^{J^*}_{j^*=1}
\phi_{ce}^{max,j^*}
\end{split}
\end{equation}
As we will exemplify in the applications section, this has the advantage of being unaffected by the underdetermination due to non-unique maximally integrated cause and effect repertoires while still depending on whether or not a particular system subset in a state specifies a concept.

\subsection*{Unidirectional partitions}\label{sec:unidirectional_part}

At this point, we have to partition the system again to define $\Phi$. This kind of partition differs somewhat from the system decomposition presented above in that it is a unidirectional partition. The aim behind unidirectional partitioning is to evaluate whether a subset $X^S \subset X^d$ has both selective causes and selective effects on its complement $X^d \setminus X^S$. Intuitively, this corresponds to noising the connections from $X^S$ to $X^d \setminus X^S$ and - in an independent calculation - the connections from $X^d \setminus X^S$ to $X^S$ ("unidirectional" partition). Again, this is readily done by factorization of the  system's original joint distribution $p_{ce}(X_t,X_{t-1})$. To this end, for a subset $X^S$, we compute two subjoint distributions, $\overrightarrow{p}_{ce}\left(X^S\right)$, where we noise the input to $X^S$ (i.e. making its current state independent by factorization), and $\overleftarrow{p}_{ce}\left(X^S\right)$ where we noise the input from $X^S$ (i.e. making its past state independent by factorization):   
\begin{align}\label{eq:unidir_cut_noise_to}
\begin{split}
\overrightarrow{p}_{ce}\left(X^S\right) &:= 
\sum_{X^S_t}p(X_t,X_{t-1})\sum_{X^d_t \setminus X^S_t}\sum_{X_{t-1}}p(X_t,X_{t-1})\\
&= p(X^d_t \setminus X^S_t,X_{t-1})p(X^S_t)
\end{split}
\end{align}
and 
\begin{align}\label{eq:unidir_cut_noise_from}
\begin{split}
\overleftarrow{p}_{ce}\left(X^S\right) &:= \sum_{X^S_{t-1}}p(X_t,X_{t-1})\sum_{X^d_{t-1} \setminus X^S_{t-1}}\sum_{X_{t}}p(X_t,X_{t-1})\\
&= p(X^d_{t-1} \setminus X^S_{t-1},X_{t})p(X^S_{t-1})
\end{split}
\end{align}
Here, we implicitly take advantage of the fact that the original joint distribution encompasses two adjacent points in time and that, therefore, every variable in $X^S_t$ has its counterpart in $X^S_{t-1}$. For the two newly defined joint distributions, we repeat the above calculations for the same system state to see whether and how many of the original concepts (maximally integrated cause and effect repertoires) we can recover and if their $\phi_{ce}^{max}$ values change. For all possible subsystems of $X^d$, we then define the unidirectional partition that makes the least difference to the original constellation of concepts as the Minimum (conceptual) Information Partition (MIP). 
IIT then essentially defines the integrated conceptual information $\Phi$ as the amount of conceptual information that is lost due to the partition over the MIP. Similarly, but again avoiding the underdetermination due to non-unique maximally integrated cause and effect repertoires, we define $\Phi$ of a system being in a state based on \cref{eq:concept_info_us} as
\begin{equation}\label{eq:PHI}
\begin{split}
\Phi(X^{d*}_t) := \sum^{J^*}_{j^*=1}
\phi_{ce}^{max,j^*} - \sum^{J^*}_{j^*=1}
\phi_{ce, MIP}^{max,j^*}.
\end{split}
\end{equation}

\subsection*{Maximally integrated conceptual information $\Phi^{max}$}\label{sec:PHI_max}

Defining the maximally integrated conceptual information $\Phi^{max}$ of a system being in a specific state corresponds to the reiteration of the above evaluation over all possible subsystems. First, there is an important conceptual distinction to make. Until this point, we have always considered a sub\textit{set} $X^S \subset X^d$ describing a set of $d$ physical elements. A sub\textit{system} $Y^b$ with $b < d$ now refers to the notion of treating $Y^b$ as a new system while regarding the elements $X^d \setminus Y^b$ as external background conditions. Formally, this corresponds to keeping the state of the outside elements fixed in the marginal conditional distribution in \cref{eq:CI}. We thus essentially define a new forward TPM over the subsystem $Y^b$ and therefore a new joint distribution based on \cref{eq:pce}. We then determine $\Phi$ as in \cref{eq:PHI} over the subsystem. This process is repeated for all possible subsystems, with the constraint that $b > 2$ because one-element subsystems cannot be partitioned and therefore cannot be integrated by definition. The maximum value of $\Phi$ over all subsystems is then defined as maximally integrated conceptual information $\Phi^{max}$ (and the corresponding subsystem is called a "complex" in IIT). Notably, IIT claims that $\Phi^{max}$ is identical to the degree to which a physical system is conscious.  

\noindent In summary, the measure of integrated information $\Phi$ rests on a standard probabilistic model approach to dynamical systems - a multivariate stochastic process that fulfils the Markov property (cf.  \cref{eq:SP,eq:MP,eq:CI}). Against this background, the integrated information of a system state is defined by the irreducibility of its conceptual structure as assessed by partitioning the system, which corresponds to the removal of stochastic dependencies between the random entities describing the system. 
The system's joint probability distribution over two adjacent points in time is uniquely defined by the system's transition probability distribution and is sufficient for all necessary mathematical operations in the evaluation of $\Phi$. 
In the following sections, we show how this general definition of $\Phi$ can be applied in the context of a specific example system.

\color{black}

\section{Results: Computing $\Phi$}\label{sec:Applications}

In the current section, we consider a concrete application of the general formulation above in a system with discrete state space which is defined non-parametrically by the explicit definition of the transition probability distribution factors as logical operations. This system corresponds to the exemplary system discussed in \citep{Oizumi2014} and serves the validation of our formulation and the illustration of quale underdetermination.

\subsection{Characterization of the system by its joint distribution}\label{sec:DSS}

\begin{figure}[htp!]
\includegraphics[width=\textwidth]{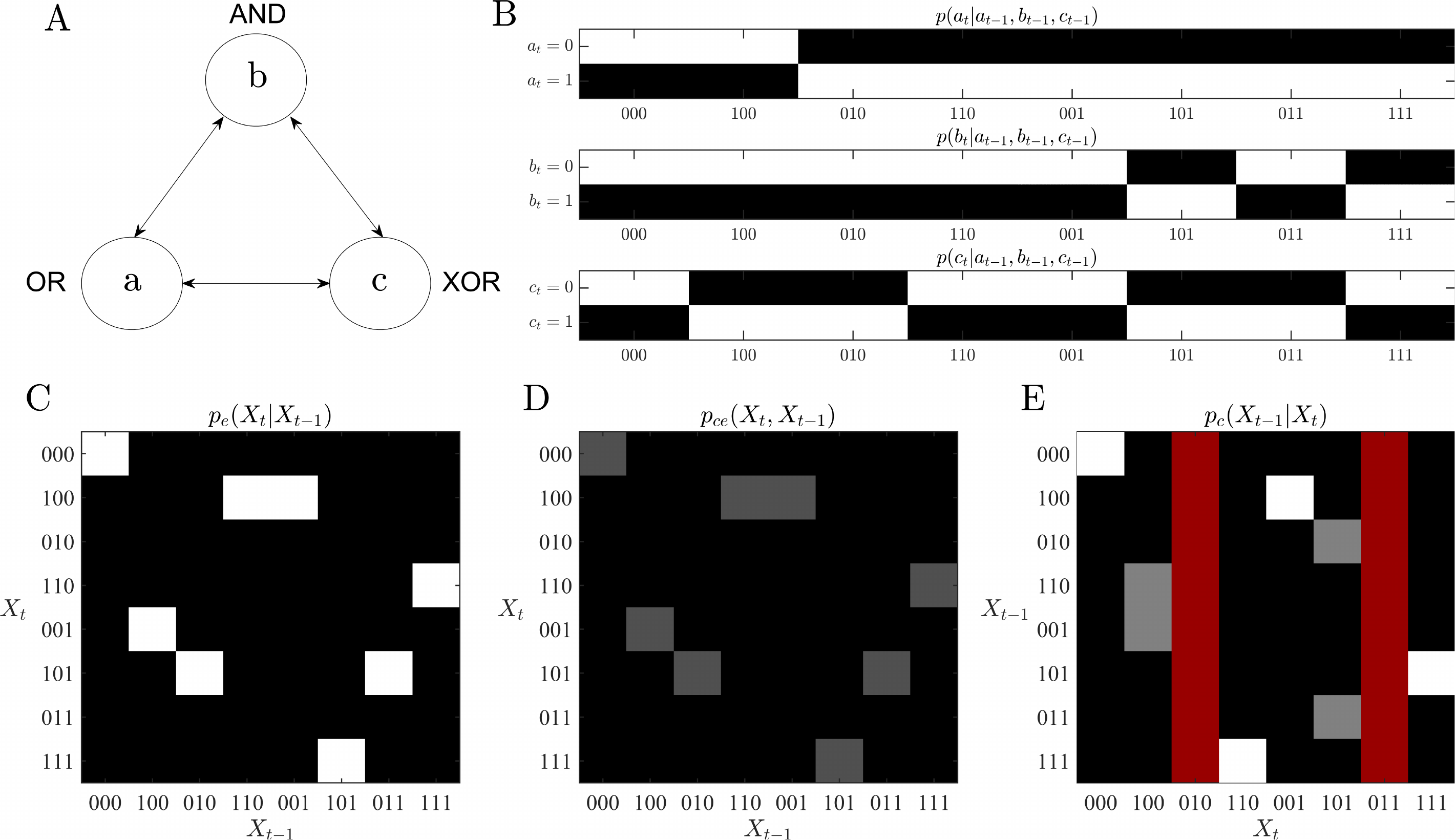}
\caption{\textbf{Characterization of the exemplary discrete state system}. The system is identical to that presented in \citet{Oizumi2014} (e.g., Figures 1 and 4 therein). Panel A shows the system comprising three random variables that implement the logical operations OR, AND, and XOR. Panel B visualizes the corresponding marginal conditional probability distributions, with black tiles indicating a probability mass of $0$ and white tiles indicating a probability mass of $1$. The product of these marginal conditional probability distributions yields the conditional distribution $p(X_t|X_{t-1})$ depicted in panel C, i.e. the state transition probability matrix. By multiplication with a maximally uncertain distribution over past states, i.e. $p_u(X_{t-1})$, the joint distribution $p_{ce}(X_t,X_{t-1})$ of panel D is obtained. Here, dark gray tiles indicate a probability mass of $0.125$. For the current example, $p_u(X_{t-1})$ corresponds to the uniform distribution over past system states. Based on the formulation presented herein, the joint distribution in panel D sufficiently characterizes the system for the derivation of $\Phi$. Moreover, conditioning $p_{ce}(X_t,X_{t-1})$ on $X_t$ yields the backward TPM $p_c(X_{t-1}|X_t)$ shown in panel E. Here, white tiles indicate a probability mass of $1$, gray tiles a probability mass of $0.5$, and red tiles represent undefined entries. These entries correspond to states of $X_t$ that cannot have been caused by any of the states of $X_{t-1}$ due to the logical structure of the network.}\label{fig:Figure_2}
\end{figure}

In discrete state systems, the random variables that model the system's elements take on a finite number of states with a certain probability mass. As an exemplary discrete state system, we consider a system presented in \citet{Oizumi2014}. This system is three-dimensional, and, in concordance with \citet{Oizumi2014}, we denote its state vector by $X_t = (a_t, b_t, c_t)$ (Figure \ref{fig:Figure_2}A). The system is defined in terms of the marginal conditional distributions of its component variables (cf. \cref{eq:CI}). Specifically, the variables  $a_t, b_t$ and $c_t$ may take on values in $\{0,1\}$, such that the outcome space $\mathcal{X}$ is defined as $\{0,1\}^3$, and implement logical operations on the state of their predecessors $a_{t-1}, b_{t-1}$ and $c_{t-1}$. As shown in Figure \ref{fig:Figure_2}B, $a_t$ implements a logical OR, $b_t$ implements a logical AND, and $c_t$ implements a logical XOR operation. Note that in this case, the relevant distributions of \cref{eq:MP} correspond to probability mass functions, which can be represented on the implementational level by high-dimensional numerical arrays.\\
 
As discussed above, the forward transition probability matrix $p_e(X_t|X_{t-1})$ of the system corresponds to the product of the marginal conditional distributions (cf. \cref{eq:CI}). This distribution is shown in Figure \ref{fig:Figure_2}C. The joint distribution $p_{ce}(X_{t-1},X_t)$ is derived by multiplication of the transition probability with a maximally uncertain distribution over past states $p_u(X_{t-1})$ (cf.\cref{eq:pce}). In this example, the maximally uncertain distribution is given by the uniform distribution over past states, i.e. $p_u(X_t = X_t^*):= |\{0,1\}^3|^{-1}$ for all $X_t^* \in \{0,1\}^3$ (cf. Figure \ref{fig:Figure_2}D). From the ensuing joint distribution $p_{ce}(X_{t-1},X_t) = p(a_{t-1}, b_{t-1},c_{t-1},a_t, b_t, c_t)$, the backward TPM $p_c(X_{t-1}|X_t)$ can be evaluated by conditioning on $X_t$. The resulting distribution is shown in Figure \ref{fig:Figure_2}E. Note that there are some undefined entries (displayed in red). These undefined entries correspond to system states that cannot have been caused by any previous state due to the constraints placed by the logical operations of the system variables. In the following, we illustrate the application of the theoretical formulation above in the evaluation of $\Phi$ for the system state $X_t = (a_t = 1,b_t = 0, c_t = 0)$. 

\color{black}

\subsection{Exclusion principle and computation of $\phi^{max}_{ce}$}
\begin{figure}[htp!]
\includegraphics[width=\textwidth]{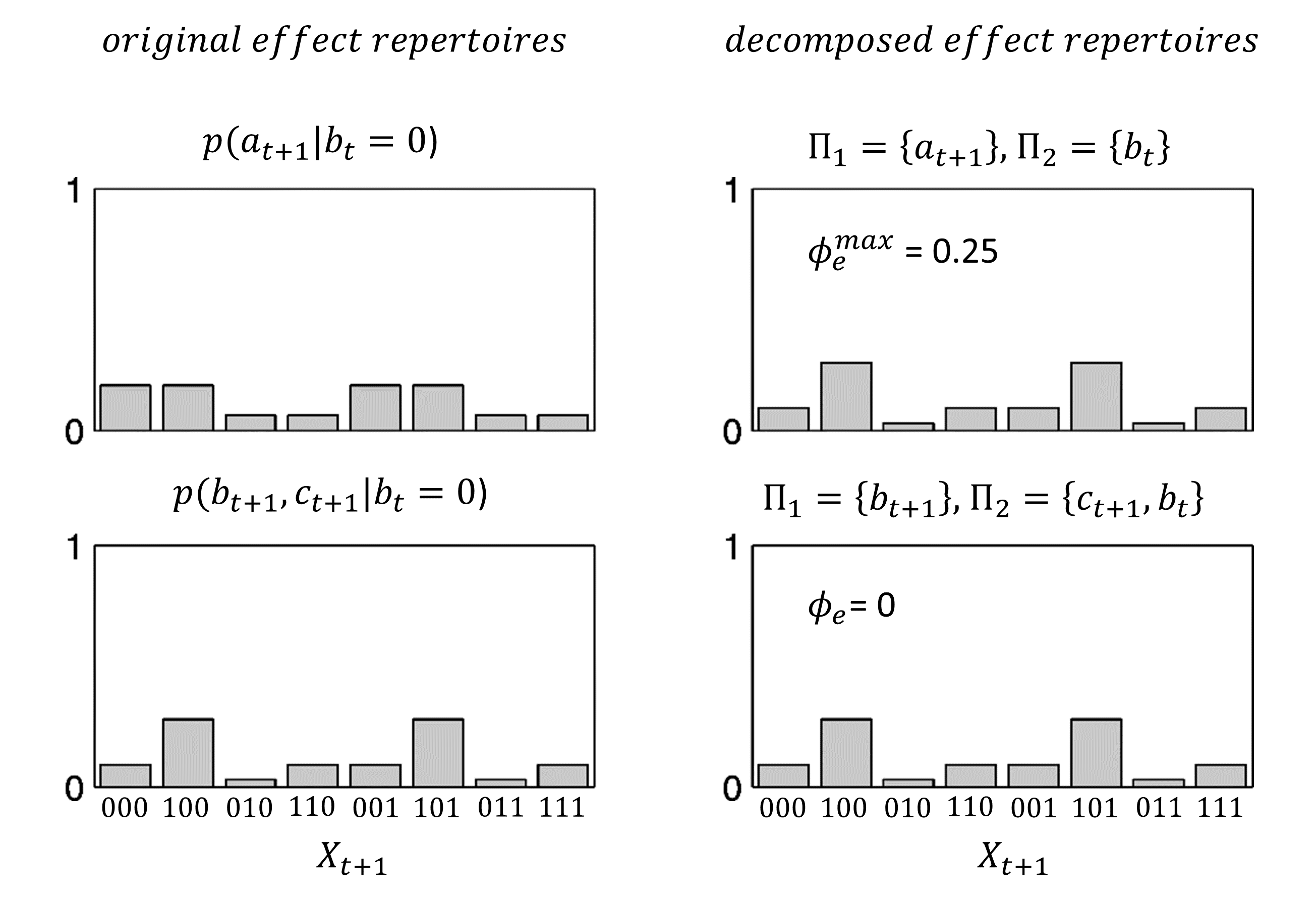}
\caption{\textbf{Exclusion principle. Evaluation of maximally integrated effect information $\phi^{max}_e$ for $b_t = 0$}. \textcolor{black}{Illustration of the computations necessary for the evaluation of maximally integrated effect information for the system subset $X^S_t = b_t$ in the state $b_t = 0$. The maximally irreducible effect repertoire corresponds to the conditional distribution $p(a_{t+1}|b_t = 0)$ with $\phi^{max}_e = 0.25$, while all other conditional distributions of the system's powerset are identically recovered by their respective minimum information partitions. The lower panels depicts one of these conditionals, $p(b_{t+1},c_{t+1}| b_t = 0)$. Note that the corresponding distributions are expanded to the whole system's states $X_{t+1}$ in order to compare conditional distributions of differing dimensionality. This is done by multiplication of the particular conditional with the marginal distribution over the respective complement with respect to $X_{t+1}$, i.e. $p(b_{t+1},c_{t+1})$ for the upper panels and $p(a_{t+1})$ for the lower panels. For the cause repertoires, this is done in analogy for $X_{t-1}$.}}\label{fig:Figure_exclusion}
\end{figure}
First, we illustrate the computation of maximally integrated cause-effect information $\phi^{max}_{ce}$ (i.e. the implementation of the exclusion principle) in the discrete state system. To this end, we focus on the example of the system subset $X^S_t = b_t$ and evaluate the maximally integrated effect information $\phi^{max}_{e}$ for this subset being in the state $b_t = 0$. Recall that this corresponds to computing the $\phi_{e}$ values for all possible conditional distributions over the system's powerset (\cref{eq:powerset_effect}), which in the example system is given by 
\begin{equation}\label{eq:powerset_example}
\mathcal{P}(X) = \lbrace \{ a \},\{ b \},\{ c \},\{ a,b \},\{ a,c \},\{ b,c \},\{ a, b, c \} \rbrace 
\end{equation} according to \cref{eq:powerset_X}. Note that the powerset is of cardinality $C = 2^3 - 1 = 7$ (cf. \cref{eq:powerset_X} and ff.). We thus compute the seven conditionals $p_{e}(X^{\mathcal{P}_j}_{t+1}| b_t = 0)$ for $j = 1,2,...,7$ to find the one whose decomposition yields the maximum $\phi_{e}$ value compared to all the others. Explicitly, we thus compute $p(a_{t+1}|b_t=0), p(b_{t+1}|b_t=0),...,p(a_{t+1},b_{t+1},c_{t+1}|b_t=0)$ and their respective decomposed variants according to the system decomposition rule (cf. \cref{eq:bpce}) and calculate the corresponding $\phi_{e}$ values based on \cref{eq:phie} (for a detailed illustration of how a single $\phi$ value is computed, the reader is referred to the Supplementary Materials section). Figure \ref{fig:Figure_exclusion} shows two out of the seven conditionals together with their decomposed variants. Note that the respective conditional distributions are always expanded to the states over the whole system (here, $X_{t+1}$) in order compare conditional distributions of differing dimensionality (see figure caption). The distribution $p_{e}(X^{\mathcal{P}_1}_{t+1} = a_{t+1}| b_t = 0)$ yields $\phi_{e} = 0.25$ over its minimum information partition (MIP) $\Pi_1 = \{a_{t+1}\}, \Pi_2 = \{b_{t}\}$. The conditional distribution $p_{e}(X^{\mathcal{P}_6}_{t+1} = b_{t+1},c_{t+1}| b_t = 0)$ on the other hand is identically recovered over its MIP, and thus $\phi_{e} = 0$. This is also true for all other five conditional distributions, so that $\phi^{max}_{e} = 0.25$, and the corresponding maximally irreducible effect repertoire is $p_e(a_{t+1}|b_t = 0)$. We proceed in analogy with the set of seven cause repertoires for $b_t = 0$ to define $\phi^{max}_{c}$. The minimum of $\phi^{max}_{c}$ and $\phi^{max}_{e}$ then defines maximally integrated cause-effect information $\phi^{max}_{ce}$ (cf. \cref{eq:max_cei}).

\subsection{Composition principle and conceptual information}\label{sec:DSS_composition}
\begin{figure}[htp!]
\includegraphics[width=\textwidth]{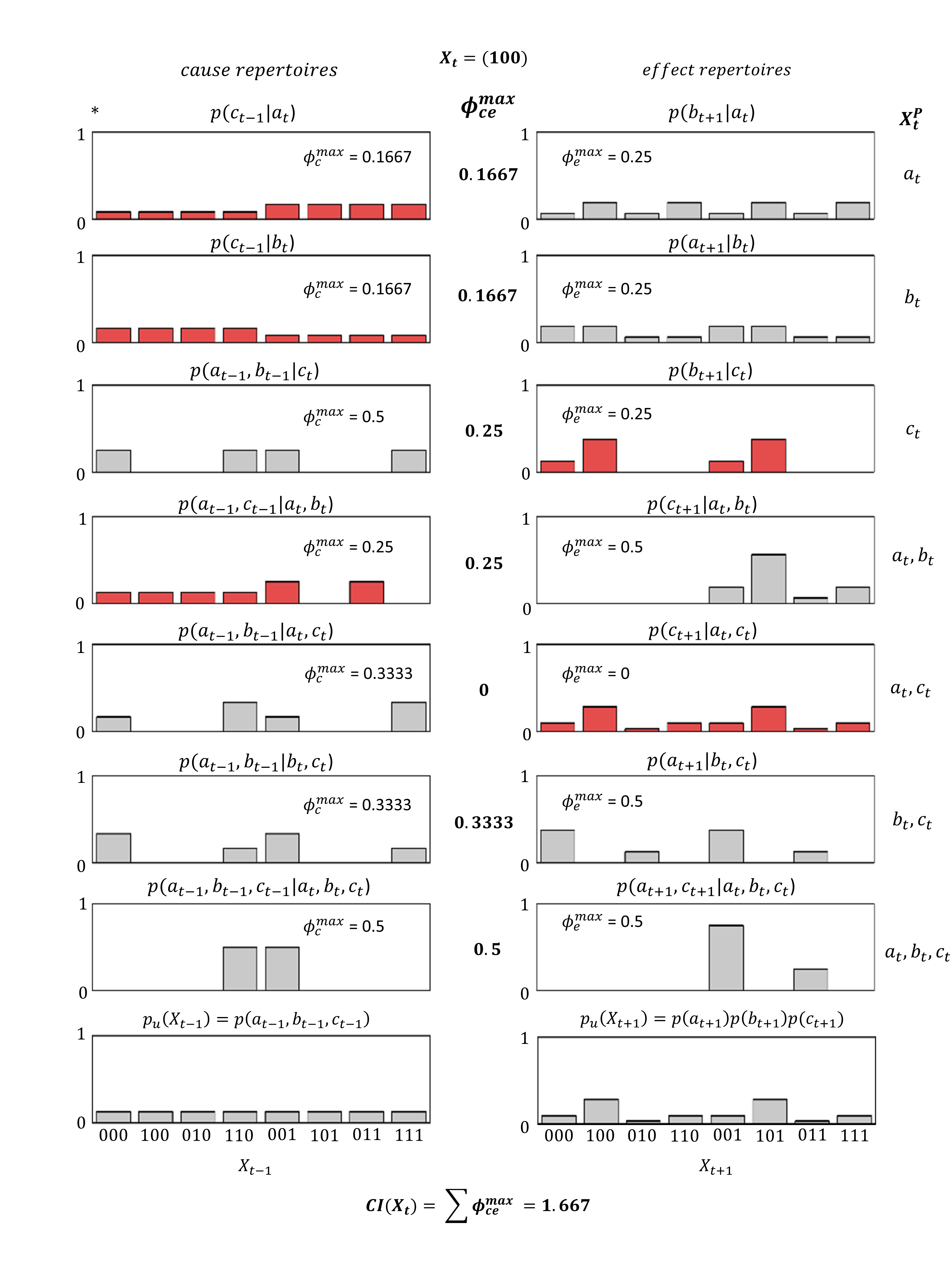}
\caption{\textbf{The set of maximally irreducible cause and effect repertoires for the system state $X_t = (1,0,0)$}. \textcolor{black}{The figure visualizes the the implementation of the composition principle, i.e. the computation of the maximally integrated cause and effect repertoires for every subset $X^\mathcal{P}_t$ in the powerset of the system elements. All those subsets for which $\phi^{max}_{ce} > 0$ form a "concept", a maximally integrated cause-effect repertoire. As we discuss in the main text, however, the distributions highlighted in red are not unique. Those distributions differ from the ones reported in \citep{Oizumi2014} as we enforce lower distribution dimensionality in underdetermined cases. The definition of conceptual information $CI$ as the sum over all $\phi^{max}_{ce}$ applied here is unaffected by non-unique repertoires. The bottom panels show the maximum entropy distributions in the respective temporal direction past ($p_u(X_{t-1})$) and future ($p_u(X_{t+1})$).}}\label{fig:Figure_concepts}
\end{figure}

To implement the composition principle, we now apply the process illustrated above not only to the subset $X^S_t = b_t$ but to all possible subsets, i.e. again over the system's power set in \cref{eq:powerset_example} according to \cref{eq:powerset_composition_cause} and \cref{eq:powerset_composition_effect}. Figure \ref{fig:Figure_concepts} visualizes the results of these calculations (similar to figs. 10 and 11 in \citep{Oizumi2014}). Based on the powerset, we thus obtain seven $\phi^{max}_{ce}$ values, one for every element in the powerset. All those elements $X^\mathcal{P}_t$ of the powerset that yield a $\phi^{max}_{ce}>0$ form a maximally irreducible cause-effect repertoire, called a "concept" in \citep{Oizumi2014}. We see that this is the case for all $X^\mathcal{P}_t$ except for $X^\mathcal{P}_t = \{a_t, c_t\}$ because the effect repertoires over this variable subset are not maximally integrated, i.e. all possible effect repertoires for $a_t = 1, c_t = 0$ yield $\phi_{e} = 0$. The example system being in the state $a_t = 1, b_t = 0,c_t = 0$ thus specifies a total of six concepts with their corresponding $\phi^{max}_{ce}$ values, which, importantly, are identical to the ones reported in \citep{Oizumi2014}. Note, however, that not all of the depicted distributions are the same as in IIT 3.0. This is because all those distributions highlighted in red correspond to cases in which the maximally integrated cause or effect repertoire is not unique, i.e. there are several conditional distributions for the particular subset $X^\mathcal{P}_t$ which yield the same maximal $\phi$ value. Note first that in case of the effect repertoires over $X^\mathcal{P}_t = \{a_t, c_t\}$, this is a logical necessity. If any of the possible conditional distributions were to specify a $\phi_{e}>0$, then that distribution would automatically become the maximally integrated effect repertoire, or, more generally, if $\phi^{max}_{c} = 0$ or $\phi^{max}_{e} = 0$, then the corresponding set of repertoires is never unique. As we can see in fig. \ref{fig:Figure_concepts}, however, there are also cases in which $\phi_{c}>0$ or $\phi_{e}>0$ and the corresponding repertoire is not unique. These cases have several important implications for IIT, which we consider to some detail in the example below.           

\subsection{On non-unique maximally irreducible cause and effect repertoires}\label{sec:non_unique_rep}
\begin{figure}[htp!]
\includegraphics[width=\textwidth]{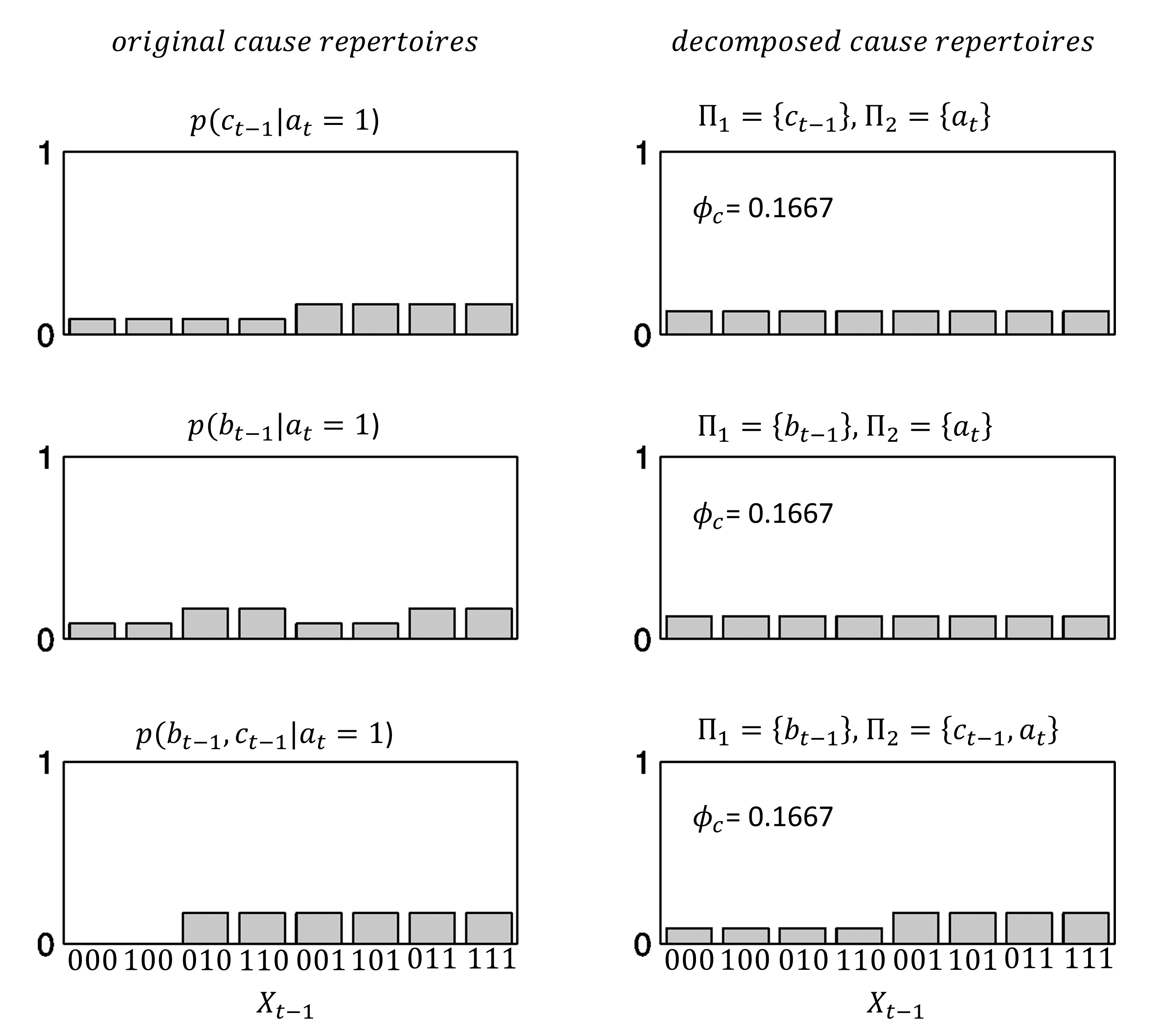}
\caption{\textbf{Non-unique maximally integrated cause repertoires over $a_t=1$}. \textcolor{black}{All three conditional distributions depicted here lead to the same maximal value of integrated cause information over their respective minimum information partitions. The top panel corresponds to the distribution shown in fig. \ref{fig:Figure_concepts}, while the bottom panel corresponds to the distribution reported in \citep{Oizumi2014}. In these cases, it is underdetermined which distribution to choose. However, the exclusion principle demands that causes should not be multiplied beyond necessity. We thus argue that exclusion favors the lower-dimensional distributions in these cases, i.e. the ones over fewer causes (cf. discussion).}}\label{fig:Figure_nonunique}
\end{figure}

In the following, we will briefly focus on the reason why non-unique maximally irreducible cause and effect repertoires are of interest to the IIT framework. First, note that the original definition of conceptual information $CI$ (\cref{eq:concept_info} and \citep{Oizumi2014}) and integrated conceptual information $\Phi$ rests on the distance between the respective maximally integrated repertoire and the maximum entropy distribution in the respective direction past or future. These distributions are depicted in the bottom panels in fig. \ref{fig:Figure_concepts}. Due to the definitions in \citep{Oizumi2014}, the values of $CI$ and $\Phi$ are thus not only dependent on the maximally integrated cause-effect information $\phi^{max}_{ce}$ but also on the actual distributions yielding these $\phi^{max}_{ce}$ values (cf. \cref{eq:concept_info}). In the case of the highlighted distributions in fig. \ref{fig:Figure_concepts}, however, there are multiple of these maximally irreducible distributions so it is underdetermined which one to choose. As an example, consider the cause repertoire over the system subset $X^S_t = a_t$ (top left panel indicated by an asterisk in fig \ref{fig:Figure_concepts}). In this case, there are in fact three distributions whose decomposition leads to the maximal value of $\phi^{max}_{c} = 0.1667$, which we visualize together with their respective decompositions in fig. \ref{fig:Figure_nonunique}. The distribution $p(c_{t-1}|a_t = 1)$ in the top panel corresponds to the one depicted in fig \ref{fig:Figure_concepts}, and the bottom panel relating the distribution $p(b_{t-1}, c_{t-1}|a_t = 1)$ is identical to the one reported in \citep{Oizumi2014}. As a side note, first consider the decomposition of $p(c_{t-1}|a_t = 1)$, which is given by the minimum information partition $\Pi_1 = \{ c_{t-1}\}, \Pi_2 = \{ a_{t}\}$. In IIT 3.0, this corresponds to the conditional $p(c_{t-1}|[])p([]|a_{t})$. Due to our decomposition rule in \cref{eq:fact_part_joint}, however, we have $z_1 = 0$ and $u_2 = 0$, and thus the decomposed cause repertoire is given by $p(c_{t-1}|a_t) = \frac{p(c_{t-1})p(a_{t})}{p(a_{t})}$. Our formulation thus eschews empty conditionals and also shows that IIT's assumption that $p(x|[]) = p(x)$ and $p([]|x) = 1$ directly follows from it. In any case, we can see from fig. \ref{fig:Figure_nonunique} that the respective partitions all yield the same $\phi^{max}_{c}$ value. In contrast, the Earth Mover's Distance to the maximum entropy distribution in the past (i.e. the uniform distribution $p_u(X_{t-1})$, see fig. \ref{fig:Figure_concepts}) may of course differ, depending on which distribution we choose. For $p(c_{t-1}|a_t = 1)$ and $p(b_{t-1}|a_t = 1)$, this evaluates to $D = 0.1667$, while for $p(b_{t-1}, c_{t-1}|a_t = 1)$, $D = 0.3333$. Since this distance measure directly contributes to the definition of conceptual information in \citep{Oizumi2014}, $CI$ and $\Phi$ can change depending on which distribution we label the maximally integrated cause repertoire. Note that the definitions of $CI$ and $\Phi$ we propose in \cref{eq:concept_info_us} and \cref{eq:PHI} are not sensitive to the actual distributions but only depend on the value of $\phi^{max}_c$ and thus we report them here. Further, IIT interprets the maximally irreducible cause-effect repertoire as a "point" in qualia space. If this repertoire is underdetermined, however, then so is the quale. It may thus be desirable to find a sensible criterion for which repertoire to choose in these cases. To this end, consider again the distributions in \cref{fig:Figure_concepts}. Here, the distribution that is reported in IIT 3.0 is of dimensionality three, while the one reported here only features two dimensions. In fact, this true for all the non-unique distributions in \cref{fig:Figure_concepts}. This is due to the fact that the computational implementation of IIT always chooses the distribution over the higher-dimensional set (the "bigger purview") because it "specifies information about more system elements" (see supplementary fig. 1 in \citet{Oizumi2014} ). In contrast to this we suggest, however, that a strict interpretation of the exclusion principle should in fact favour the lower-dimensional distributions. Recall that the exclusion principle postulates that causes and effects should not be multiplied beyond necessity. As such, choosing the distribution $p(b_{t-1}, c_{t-1}|a_t = 1)$ in fig. \ref{fig:Figure_nonunique} (over "two causes") thus seems less parsimonious than choosing one of the lower-dimensional distributions over fewer causes. Throughout the manuscript, we thus always enforce the lower dimensionality in cases of underdetermination, and return to this issue in the discussion.  

\subsection{Integrated conceptual information $\Phi$}

\begin{figure}[htp!]
\includegraphics[width=\textwidth]{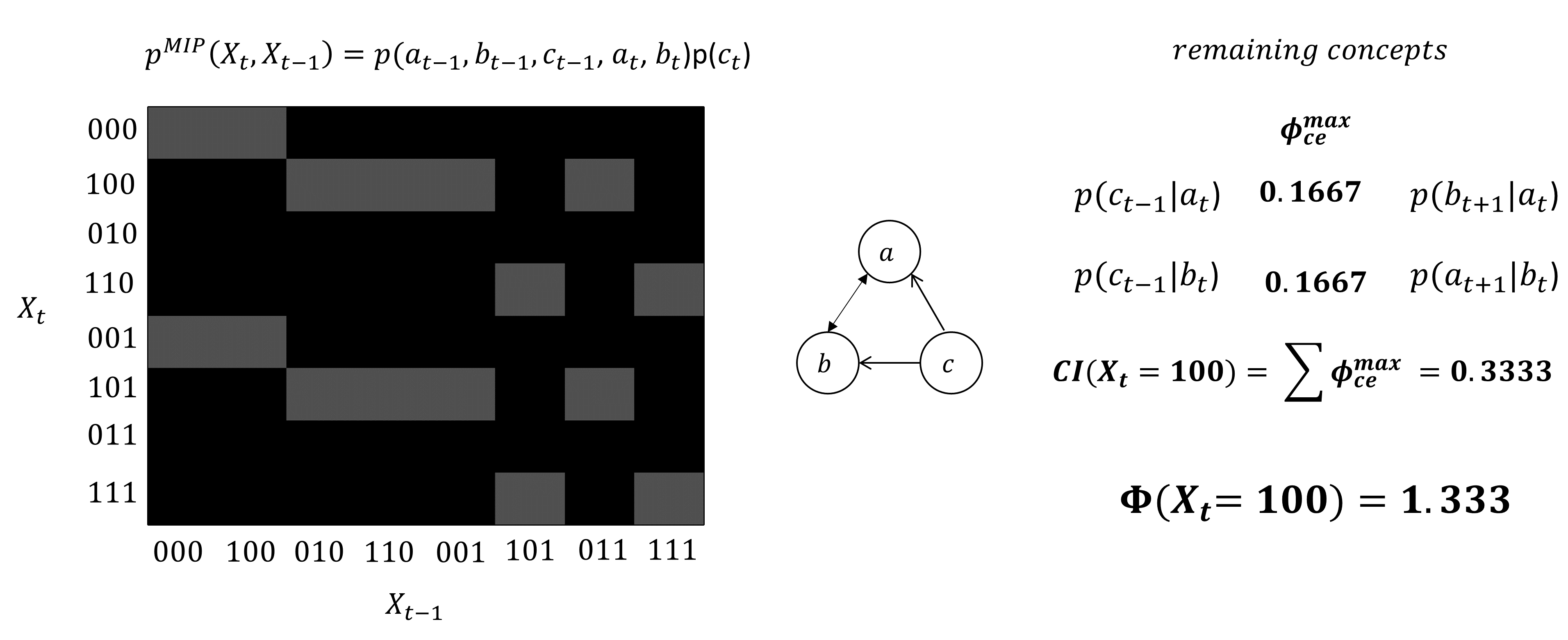}
\caption{\textbf{Integrated conceptual information $\Phi$ for the discrete example system in state $X = (1,0,0)$}. \textcolor{black}{The joint distribution on the left corresponds to the (system state's) minimum information partition (MIP). Here, the MIP is given by factoring $c_t$ out of the original joint distribution in fig. \ref{fig:Figure_2}, i.e. "noising the connections" from $c$ to $a, b$ (see network depiction in the center). Gray tiles refer to a probability mass of 0.0625. The thus factorized joint distribution recovers two of the original six concepts in fig. \ref{fig:Figure_concepts}, while $\phi^{max}_{ce}$ for the remaining four reverts to zero. For the current system state, we thus find that $\Phi = 1.333$, based on \cref{eq:PHI}. }} \label{fig:Figure_PHI_X100}
\end{figure}

\noindent We can now illustrate the computation of integrated conceptual information $\Phi$ as defined by \cref{eq:PHI}. Recall that the definition of $\Phi$ requires unidirectional system partitions according to \cref{eq:unidir_cut_noise_to} and \cref{eq:unidir_cut_noise_from} in order to find the (system state's) minimum information partition. For the given example state, this evaluates to the factorization depicted in \cref{fig:Figure_PHI_X100}. The unidirectional MIP is given here by factoring out $p(c_{t})$, which corresponds to noising the connections from $c$ to $a$ and $b$. Note the difference between the thus factorized joint distribution and the original joint distribution in \ref{fig:Figure_2}. Based on this joint distribution, we thus reiterate the presented formulation and find that two out of the six original concepts are identically recovered while the other four vanish to $\phi^{max}_{ce} = 0$. The conceptual information over the unidirectionally partitioned system is thus $CI = 0.3333$ according to \cref{eq:concept_info_us}. Based on \cref{eq:PHI}, we thus obtain $\Phi = 1.333$.

\subsection{Maximally integrated conceptual information $\Phi^{max}$}

\begin{figure}[htp!]
\includegraphics[width=\textwidth]{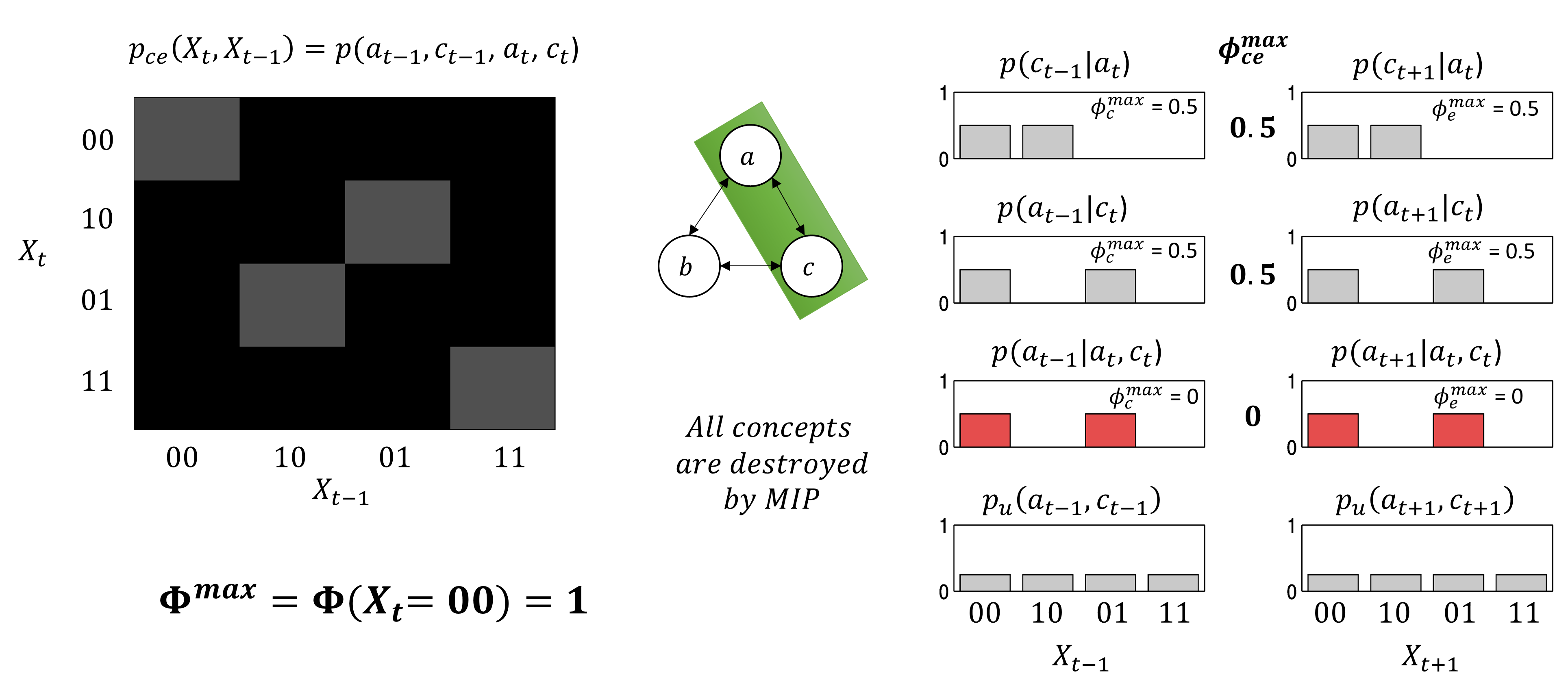}
\caption{\textbf{Maximally integrated conceptual information $\Phi^{max}$ for the discrete example system in state $X_t = (0,0,0)$}. \textcolor{black}{The maximum value of $\Phi$ over all possible subsystems is found in the subsystem $\{a, c\}$, highlighted in green in the network graph. Defining the subsystem corresponds to computing the new joint distribution on the left, which is found by keeping the state of $b = 0$ fixed in the marginal conditional distributions (fig. \ref{fig:Figure_2} and \cref{eq:CI}). Here, gray tiles indicate a probability mass of 0.25. Again, based on our formulation, the subsystem is fully characterized by this joint distribution. For the state $a_t = 0, c_t = 0$, the subsystem specifies two concepts (the distributions in red are not unique because $\phi^{max}_{c} = 0$ and $\phi^{max}_{e} = 0$). Both these concepts vanish over the MIP, yielding $\Phi^{max} = 1$.}}\label{fig:Figure_PHI_MAX_000}
\end{figure}

Finally, we briefly consider the evaluation of maximally integrated conceptual information $\Phi^{max}$. To this end, we evaluate $\Phi$ as illustrated above for every subsystem of a set of $d$ elements. Recall from section \ref{sec:PHI_max} that only subsystems with at least two elements are considered (because one-element sets cannot be partitioned and are therefore not integrated by definition) and the state of all elements outside of the subsystem are fixed. This corresponds to defining a new transition probability distribution according \cref{eq:CI} and thus a new joint distribution based on \cref{eq:pce}. For the example system, the possible subsystems are given by $\{a, b\}, \{a, c\}, \{b, c\}$, and $\{a, b, c\}$. For a system state of interest, we thus obtain four $\Phi$ values, the maximum of which yields $\Phi^{max}$. In the current example of system state $a_t = 1, b_t = 0, c_t = 0$, $\Phi^{max}$ is found over $\{a, b, c\}$ and thus corresponds to the value depicted in \cref{fig:Figure_PHI_X100}. To illustrate the above, we choose a different system state, $a_t = 0, b_t = 0, c_t = 0$, and compute $\Phi$ for each of the four subsystems. For this state, the whole system $\{a, b, c\}$ specifies four maximally irreducible cause-effect repertoires and $\Phi = 0.583$. The maximum $\Phi$ value for this system state, however, is found over the subsystem $\{a, c\}$, depicted in \cref{fig:Figure_PHI_MAX_000}. Note that for this subsystem, the state of element $b$ is fixed at $b = 0$, regardless of time. On a computational level, this is conveniently implemented by discarding all those states in which $b = 1$ from the marginal conditional distributions in \cref{fig:Figure_2}. With these new marginal conditional distributions, we then form the new forward TPM according to \cref{eq:CI} and find the joint distribution $p_ce(a_{t-1}, c_{t-1}, a_{t}, c_{t})$ with a maximum entropy distribution over past states $p_u(a_{t-1}, c_{t-1})$ based on \cref{eq:pce}. The thus specified system yields two maximally irreducible cause-effect repertoires (concepts) which vanish to $\phi^{max}_{ce} = 0$ over the MIP. Thus, we find that $\Phi^{max} = 1$.  

\section{Discussion} 

In the present work, we have developed a comprehensive general formulation of integrated information theory, starting from its most recent instantiation in \citep{Oizumi2014}. This formulation rests on a standard probabilistic modelling approach, and we argue that it provides several improvements over previous formulations. Specifically, we show that all necessary mathematical operations in the derivation of $\Phi$ are sufficiently specified by a system's joint distribution $p_{ce}(X_{t-1},X_t)$ over two adjacent points in time. We present a constructive rule for the decomposition of the system into two disjoint subset, which corresponds to flexible marginalization and factorization of this joint distribution. We increase the parsimony of IIT because, while yielding the same computational results, our formulation omits interventional calculus (and specifically system perturbations), which also allows us to suspend the use of virtual elements introduced in \citep{Oizumi2014}. In this regard, we show that virtualization \textit{is} factorization and that our approach eschews the occurrence of empty conditional distributions.     
On the implementational level, our formulation is readily applied to non-parametric discrete state systems, as validated in the exemplary system from IIT 3.0. Here, we also illustrate a previously unexplored theoretical issue, which regards the underdetermination of $\Phi$ due to non-uniqueness of maximally integrated cause and effect repertoires. We propose that a strict interpretation of the exclusion postulate should favour lower-dimensionality probability distributions in these cases, and we elaborate on this issue below. Related to this aspect is the sensitivity of $\Phi$ to qualia shape, which we account for by defining $\Phi$ merely as a function of maximum integration, regardless of which distribution is maximally integrated. Ultimately, the aim is to evaluate $\Phi$ in actual empirical data. The formulation of information integration presented herein is comprehensive and general enough to pave the road towards a transfer of the IIT framework to the realm of functional neuroimaging data in the future. While system perturbations may be cumbersome in continuous systems, the estimation of parameterized joint distributions from data is common practice in other fields of empirical neuroscience, rendering it the more promising approach.  
Contrasting our formulation with similar endeavors, we now turn to some open questions as regards IIT, focusing on the boundedness of $\Phi$, system partitions, the underdetermination issue, and the state-dependency of integrated information.\\    

\color{black}

In similar spirit to the present work, \citet{Barrett2011} were among the first to propose a measure of integrated information that allows for the evaluation of integration in continuous time-series data. The main difference to the formulation presented here is that Barrett and Seth focus on an earlier version of the theory - integrated information theory 2.0 \citep{Balduzzi2008} - which is based on entropy-related measures, while we start explicitly from the most recent instantiation of IIT. As has been pointed out by \citet{Tegmark2016} and \citet{Oizumi2016}, the integrated information measure proposed by \citet{Barrett2011} can become negative. To some researchers in the field, this somewhat complicates its interpretation, although the interesting question has been raised of whether ``negative integration'' could reflect redundancy in a system \citep{Barrett2015}. Note that $\phi_{ce}$ as presented herein is bounded by zero because the EMD cannot be negative \citep{Levina2001,cover2012elements}. \color{black} $\Phi$ as given in \cref{eq:PHI} is generally expected to lie in the interval between zero (if there is a unidirectional partition that identically recovers the concepts) and the conceptual information of the unpartitioned system (if all concepts are destroyed by the minimum information partition) and could become negative if and only if the conceptual information of the partitioned system is actually greater than that of the unpartitioned system. This is counter-intuitive, of course, because it would mean that we somehow generate information by cutting the system. On a subtle note, however, the online documentation of the PyPhi code (\url{http://pyphi.readthedocs.io}) states that in rare cases this can actually occur, referred to as "magic cuts". As we will elaborate below, the issue then lies not so much in the definition of $\Phi$, but in the current definition of the system's minimum information partition. 
\color{black}
Introducing theoretical requirements for the lower and upper bounds for integrated information was also one of the main motivations for recent work by \citet{Oizumi2016}, who moreover evaluate their measure of information integration in monkey electrocorticogram data. The major differences to our work are that the authors develop their measure based on IIT 2.0 and that it involves ``atomic partitioning'', which we discuss in more detail below. \color{black}A detailed study of the boundedness of integrated information and its relation to state differentiation is found in \citep{marshall2016integrated}.   

An impressive collection of different mathematical options to measure information integration was presented recently by \citet{Tegmark2016}. Similarities between our approach and that of \citet{Tegmark2016} include the application of IIT to Markov process system models and the ensuing construction of cause and effect repertoires. We differ, however, in the constructive algorithm for the system decomposition by factorization and in that we present a comprehensive formulation up to $\Phi^{max}$. Importantly, we consider all possible partitions in the system decomposition, both symmetrical and asymmetrical. Furthermore, Tegmark does not arrive at a definition of integrated information $\Phi$ on a system level. In fact, the statement that "$\Phi$ is the minimum of $\phi$ over the exponentially many ways of splitting the system into two parts" is actually closer to the definition of $\phi_{ce}$ and thus differs substantially from IIT's definition of $\Phi$ as integrated conceptual information. However, the taxonomy for measuring choices (especially in the realm of distance measures) and using graph theory-based approximations for speeding up the necessary computations is certainly a very promising approach in the further development of IIT. 

\subsection{On partitions and boundedness}

The concept of information integration rests on the general idea that the whole is more than the sum of its parts. As such, partitioning a system is a key aspect of IIT. First, it is useful to highlight again a subtle but important distinction. The system decomposition presented in sec. \ref{sec:SysDecomp} corresponds to bipartitioning the set of random variables in order to compute a particular value of integrated cause-effect information $\phi_{ce}$, while the unidirectional (system) partitions presented in sec. \ref{sec:unidirectional_part} yield the integrated conceptual information $\Phi$ (over many individual evaluations of $\phi_{ce}$). As we have shown above, a flexible factorization of the system's joint distribution parsimoniously yields both types of partitions. In the following, we first focus on the former and the ensuing numerical issues and then turn to the latter and some conceptual issues.\\ \color{black}
As shown in the Appendix, the number of unique bipartitions of any set with cardinality $n$ is given by an identity of the Stirling number of the 2$^{nd}$ kind. The problem is that this number grows exponentially with $n$, thus seriously compromising the computational tractability for sets containing more than about $15-20$ dimensions. There have been several approaches to this problem. \citet{Tegmark2016} limits himself to symmetrical bipartitions. However, most bipartitions in a large set of elements are in fact symmetrical. This approach thus concedes generality but does not gain much computational tractability. It is furthermore unclear how this will fare in networks containing an odd number of elements as these have no strictly symmetrical bipartitions. 
\citet{arsiwalla2016global} propose a maximum information partition as a solution to combinatorial explosion where the latter is "defined as the partition of the system into its irreducible parts", of which there is naturally only one, regardless of the network size. This corresponds to the atomic partitions (i.e. the complete factorization of conditional distributions over all variables) used by \citet{Oizumi2016} to calculate a modified measure of integrated information in primate electrocorticogram data. The authors concede that atomic partitioning and to a lesser extent also symmetrical partitioning will overestimate information integration because it tends to maximize rather than minimize the informational difference the partition makes to a set of system elements. While this may be useful approach in many cases, as regards consciousness IIT aims to describe how much the system actually integrates, not how much it can maximally integrate. Thus, if we wish to stick with the current definition of $\phi_{ce}$, the question still remains which computational partitioning approach is best when the theoretical analysis of all bipartitions is no longer feasible. Our formulation places no prior constraint on the theoretically possible partitions in order to maintain generality and thus suffers from the same limitations of combinatorial explosion as previous endeavours. To overcome the numerical issues, there are a couple of outlooks that may be worth discussing. First, one approach could be to start evaluating $\Phi$ on a macroscopic scale, i.e. over merely a few brain regions of interest containing lots of neural elements, thus circumventing combinatorial explosion by scaling. Indeed, such large-scale ``hot zones" for the neural correlates of consciousness have been identified in posterior cortical zones in recent years \citep[e.g.][]{Koch2016}. While this disregards the recently developed concept of causal emergence on different spatio-temporal scales \citep{hoel2013quantifying,Tononi2016}, it is a start, and similar approaches are successfully used in dynamic causal modelling \citep{Friston2003} as well as graph-theory and mean-field-based measures \citep{Deco2015} that try to capture the behaviour of large-scale brain networks.
A more formal approach could be to find an estimate of which partitions are likely to result in a great difference to the respective repertoire over the unpartitioned set of system elements and then discard these (because $\phi_{ce}$ is defined as the minimum over these differences). For example, in a complex system, one could analyse the network structure based on graph theoretical measures and then reduce the number of partitions that have to be evaluated by discarding all partitions that cut through the connections of (i.e. introduce stochastic independence on) a hub node. Since hub nodes are by definition strongly connected and can thus be assumed to ``make a difference to the system" \citep{Oizumi2014}, such an approach could both substantially reduce the computational load (because many partitions would affect the hub) and serve as a theoretically and biologically plausible approximation.\\ 
\color{black} As regards the unidirectional partitions, there is an additional functional issue with respect to the minimum information partition on a system level. As we have addressed above, the current definition of the unidirectional partitions can yield so-called "magic cuts", which, formally, correspond to the emergence of maximally integrated cause-effect repertoires induced by the system partition. While this is not the case in the results presented herein, it raises general concerns regarding the current definition of the system partitions. Recall that with unidirectional partitioning, we are looking for the \textit{minimum} information partition, i.e. the one that makes the least difference to the original system. The emergence of previously absent concepts due to a particular partition should therefore strongly argue against that partition being regarded as the MIP because it obviously makes a profound difference to the unpartitioned system. Note that "magic cuts" also violate the very basic intuition behind the theory, namely that the whole is - functionally - more than the sum of its parts, because in some cases the sum of its parts can in principle be more than the whole.   

The above essentially amounts to the general question of whether the MIP should be defined based on state-space (i.e. the difference it makes to the set of maximally integrated cause-effect repertoires) or integration (i.e. the difference it makes to the conceptual information) or perhaps a combination of both. The latter corresponds to the idea that the original system should be an upper bound on the partitioned system over the MIP in both a qualitative and a quantitative sense. The emergence of new concepts due to a system partition can violate either, however, and therefore requires a closer examination in the future. In the spirit of our formulation, one desirable solution would be to find the MIP of a system state based on a functional similarity index comparing the original and unidirectionally partitioned joint distributions.\\[.5cm]   
 
\subsection{Underdetermination of maximally integrated cause-effect repertoires}
\color{black}
We now return to the issue of underdetermined concepts, i.e. maximally irreducible cause-effect repertoires (cf. \cref{sec:non_unique_rep,sec:exclusion}). Note that this issue has direct consequences on IIT's application to consciousness. If the maximally integrated cause-effect repertoires are underdetermined, then, based on the distance measures in \cref{eq:concept_info}, so are the conceptual information $CI$, the integrated conceptual information $\Phi$, and the maximally integrated conceptual information $\Phi^{max}$, which IIT postulates to be identical to the quantitative consciousness of a system in a certain state. Moreover, IIT interprets a maximally irreducible cause-effect repertoire as a "quale sensu stricto" \citep{Oizumi2014} and the particular set of concepts associated with $\Phi^{max}$ as a description of the actual phenomenological experience (a constellation in qualia space), which in turn is also underdetermined in these cases (quale underdetermination). With the formal definitions in \citep{Oizumi2014}, IIT thus combines the measure of quantitative consciousness, $\Phi^{max}$, with the measure of qualitative consciousness, the associated structure of concepts in qualia space, because the value of the former depends on the actual arguments of the latter. As the authors note themselves, however, the content of phenomenological experience is not necessarily a prerequisite for the degree of consciousness (e.g. in certain meditative practices reaching high-level awareness with low phenomenological content \citep{Oizumi2014}). In other terms, $\Phi^{max}$ should be sensitive to whether or not there is a conscious experience and not to the content of that experience. Formally, a quantitative measure of consciousness based on information integration should thus be a priori independent of "what" the system in a state integrates and only rely on "how much" the system in a state is integrated, similar to the definition in \cref{eq:PHI}.\\ 
In any case, we argue that the underdetermination is an aspect of the theory that requires further examination. As we have demonstrated in the discrete state example system, the computational implementation in IIT currently chooses the higher-dimensional repertoire in these cases. Due to the exclusion postulate that causes and effects should not be multiplied beyond necessity, however, we argue that the more parsimonious choice would in fact be the repertoire with the lowest dimensionality, i.e. over the fewest possible number of causes or, respectively, effects that are still maximally integrated. As the reader can see in the example in fig. \ref{fig:Figure_nonunique}, this criterion would discard the distribution reported in \citep{Oizumi2014} but still leaves two distributions with minimum dimensionality. In order to find a sensible criterion of which distribution to label the maximally integrated cause repertoire in this case, one approach would be to choose the distribution over those system elements that contribute most to the constellation of concepts as a whole (the ones that most "shape" the conceptual structure in the unique cases). Formally, this could for instance be evaluated by the number of unique concepts to which a particular subset contributes in the respective backward or forward temporal direction. In the case of fig. \ref{fig:Figure_nonunique}, for instance, $b_{t-1}$ contributes more to the unique cause repertoires than $c_{t-1}$ over all system subsets in the past. We would therefore choose the distribution $p(b_{t-1}|a_t = 1)$ as the maximally irreducible cause repertoire that most shapes the conceptual structure. While, in the given example, this criterion uniquely identifies the distribution we ought to choose, it is of course not guaranteed that this will always be the case, and surely further clarification of this issue in terms of a comprehensive formal criterion is required. On a phenomenological level, however, choosing the element which most contributes to the whole conceptual structure could perhaps make intuitive sense. Conscious experience features a set of distinct, yet unified phenomenological aspects, where some - such as a blaring sound or a blatant color - can seem to be in the foreground because they shape the unified experience more than other aspects which are also consciously experienced. \color{black} 

\subsection{State-dependency and temporal dynamics of $\Phi$}

Finally, we consider the state-dependency of $\Phi$. Recent work by \citet{Virmani2016} aims at bridging the gap between IIT and the perturbational complexity index, a practical measure based on TMS stimulation with simultaneous EEG recordings that has been successfully applied in the clinical quantification of consciousness \citep{Casali2013}. To this end, the authors derive a measure of compression-complexity which is calculated by a maximum entropy perturbation of each node in an atomic partition. While this is certainly a promising approach, their measure shows minimal dependency on the current system state, similar to the measure proposed by \citet{Barrett2011}. Based on empirical studies \citep{Koch2016}, however, it seems likely that any measure that quantifies consciousness \textit{should} indeed be state-dependent (cf. \citep{Tegmark2016}). On theoretical grounds, state-independence would also violate the selectivity postulate that IIT proposes as a prerequisite for information (e.g. Figure 3 in \citep{Oizumi2014}). Moreover, insensitivity to system states raises the question of whether a measure in fact represents integration or rather the capacity to integrate \citep{Barrett2011}. In the present work, state-dependency is preserved because whether or not a system specifies a set of maximally irreducible cause-effect repertoires depends on the conditional distributions over the state of its elements. 

\section*{Conclusion}
\color{black}
Integrated information theory is one of the leading theories in the study of consciousness, not least because it is arguably the first rigorous attempt at an analytical formalization of what is necessary for a physical system to have phenomenological experience.    
With the presented general and comprehensive formulation of integrated information in the language of probabilistic models, we hope to make a constructive contribution to the parsimony and formal as well as conceptual improvement of integrated information theory, in the spirit of ultimately transferring IIT to the realm of empirical evaluation. 
\color{black}

\section*{Data availability}

Custom-written Matlab code (The MathWorks, Inc., Natick, MA, United States) was used to implement the formulation of integrated information presented herein. The corresponding files are available from the corresponding author and from the Open Science Framework (\url{https://osf.io/nqqzg/}).

\bibliographystyle{apalike}
\bibliography{References_revision}{}

\vspace{1cm}

\section*{Supplementary Material}
Here we provide some detailed explanations and useful illustrations of the formulation presented herein. Specifically, appendix A features the formal derivation of the number $k$ of unique bipartitions in an $n$-dimensional set. Appendix B depicts a detailed example calculation of how a single value of integrated cause-effect information $\phi_{ce}$ is calculated for a particular subset of system variables in a particular state.
To this end, we list all $k = 31$ decomposed effect and cause repertoires for the exemplary discrete state system evaluated over the whole set of system elements and in state $X = (1,0,0)$.\\ 
Furthermore, the supplementary figures \ref{fig:Figure_effect_ex} and \ref{fig:Figure_cause_ex} exhaustively illustrate the system decomposition rule over all possible 7 bipartitions of a two-dimensional system and the ensuing effect (fig. \ref{fig:Figure_effect_ex}) and cause repertoires (fig. \ref{fig:Figure_cause_ex}). 

\begin{appendix}
\subsection*{Appendix A}\label{sec:Appendix_A}
Finding the number of partitions of a set containing a finite number of elements is a well-known problem in combinatorics \citep{Cameron1994}. The number of all  bipartitions of a finite set can be evaluated by the Stirling number of the 2$^{nd}$ kind,
\begin{equation}
\label{eq1}
S_{n,m} = \frac{1}{m!}\sum\limits_{i=0}^m {{(-1)}^{m-i}\binom{m}{i} } i^{n},
\end{equation}
which counts the number of possible partitions of a set of cardinality $n\in\mathbb{N}$ into $m\in\mathbb{N}$ non-empty disjoint subsets. For the purposes of IIT, bipartitions are required. Thus, $m = 2$ and we have 
\begin{small}
\begin{align}
\begin{split}
S\left(n,2\right) &= \frac{1}{2!}\sum\limits_{i=0}^2 {{(-1)}^{2-i}\binom{2}{i}} i^{n} \\
&=\frac{1}{2!}\left(\left(\left(-1\right)^{2-0}\binom{2}{0}0^{n}\right)+ \left(\left(-1\right)^{2-1}\binom{2}{1}1^{n}\right)+ \left(\left(-1\right)^{2-2}\binom{2}{2}2^{n}\right)\right)\\ 
&=\frac{1}{2}\left(\left(\left(-1\right)^{2}\left(1\right)0^{n}\right)+ \left(\left(-1\right)^{1}\left(2\right)1^{n}\right)+ \left(\left(-1\right)^{0}\left(1\right)2^{n}\right)\right)\\
&=\frac{1}{2}\left(0+ \left(-2\right)+ 2^{n}\right)\\
&=-1+ \frac{2^{n}}{2}\\
&=2^{n-1}-1
\end{split}
\end{align}
\end{small}
In the main text, we use the denotation $k := 2^{2d-1}-1$ for a set of cardinality $n = 2d$.


\newpage
\subsection*{Appendix B}\label{sec:Appendix_B}

\begin{figure}[htp!]
\includegraphics[width=\textwidth]{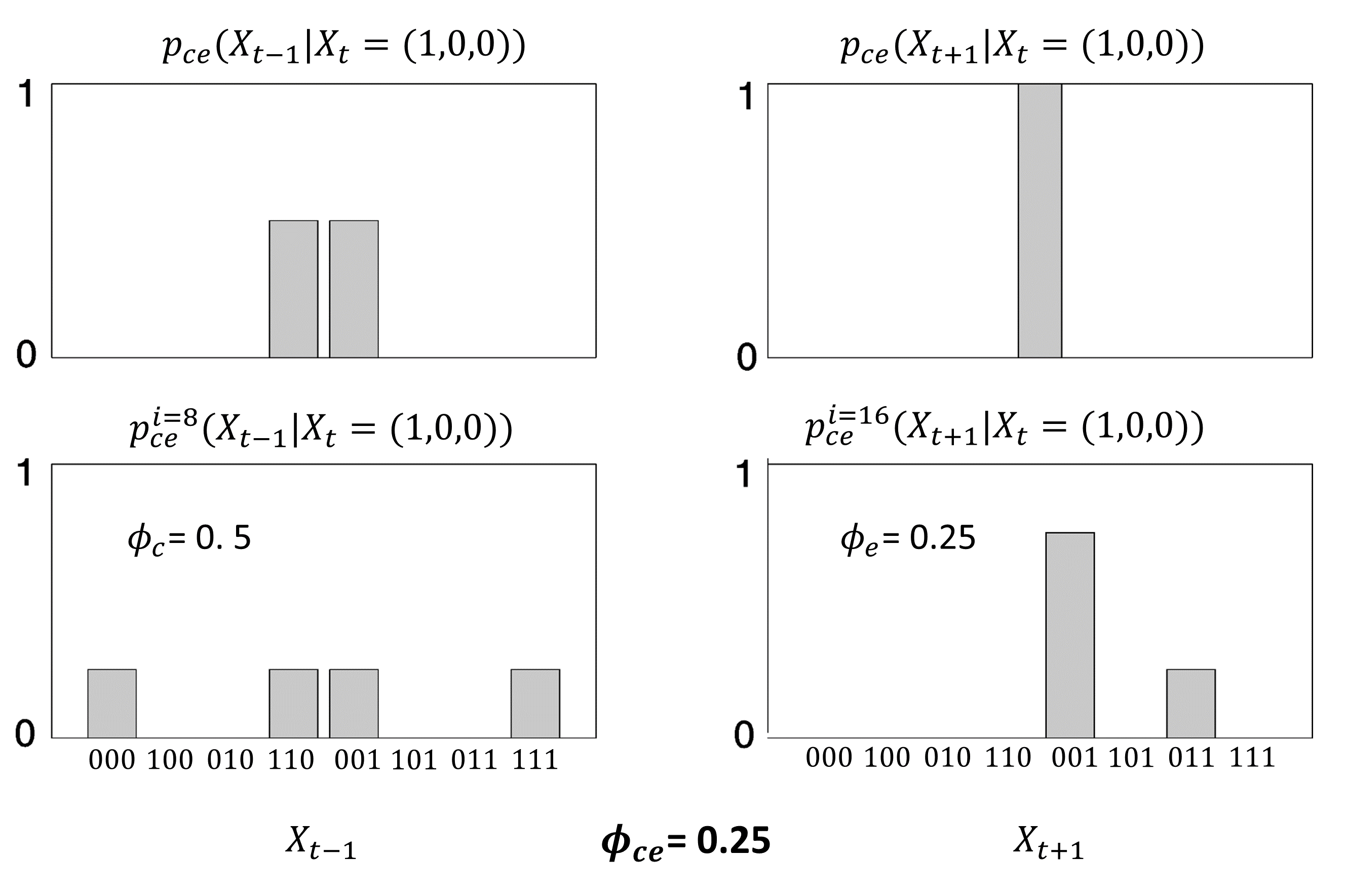}
\caption{\textbf{State-specific evaluation of $\phi_{ce}$ in the exemplary discrete state system.} The Figure depicts the probability mass functions required for the evaluation of $\phi_{ce}$ for the system state $X = (1,0,0)$ over the subset $X^S = X^d$ (i.e. the whole system) in the discrete state system introduced in the main text. The right panels visualize the relevant distributions required for the evaluation of $\phi_e((1,0,0))$, and the left panels the relevant distributions required for the evaluation of $\phi_c((1,0,0))$. The minimum of these two values is defined as the integrated cause-effect information $\phi_{ce}((1,0,0)) = 0.25$ of the system state.}\label{fig:Figure_example_phi}
\end{figure}

\begin{figure}[h!]
\includegraphics*[width=\textwidth]{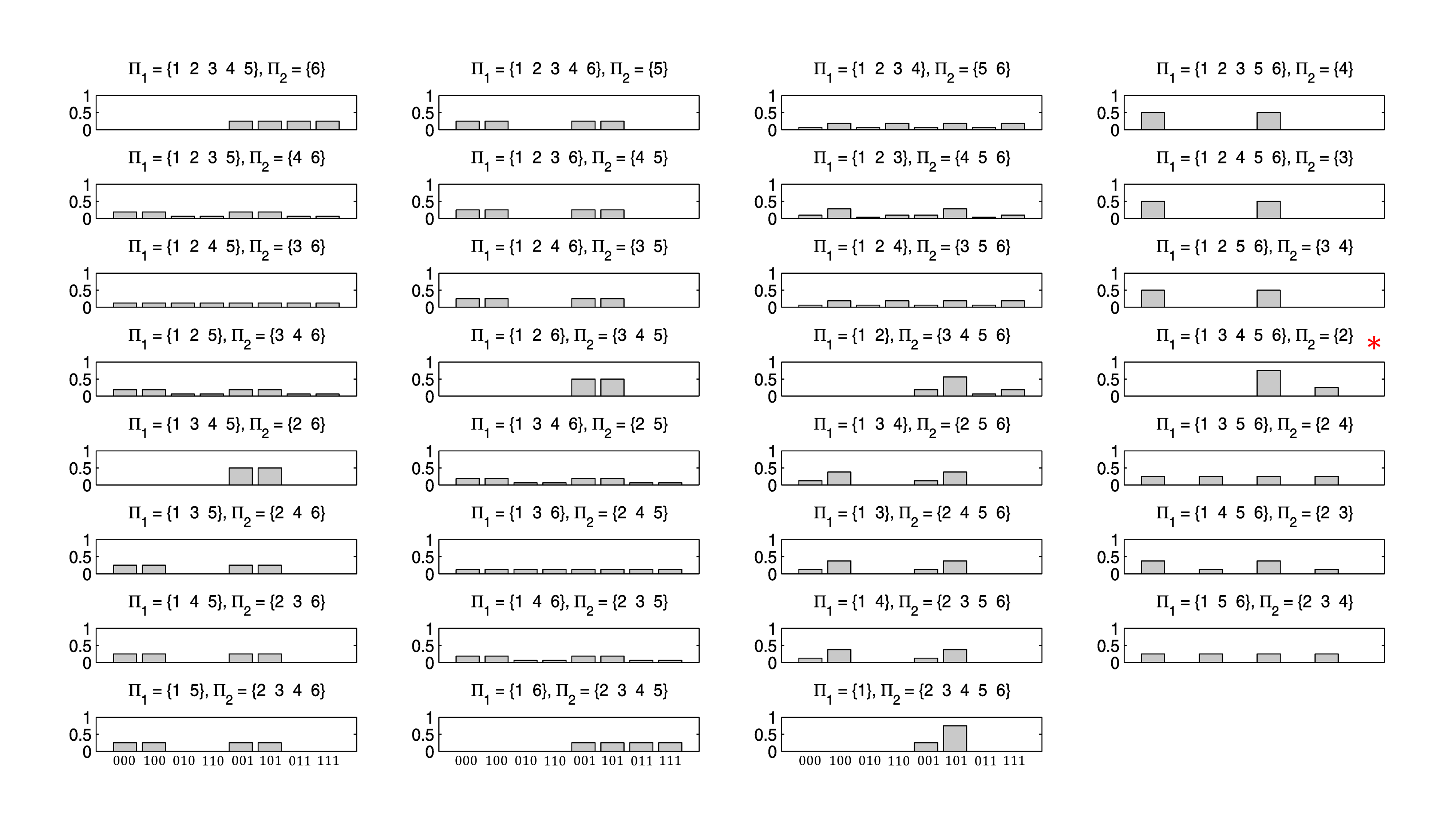}
\caption{{\bf Decomposed effect repertoire for an exemplary discrete-state system.} For the system state $X = (1,0,0)$, the figure visualizes the complete set of decomposed variants of the effect repertoire $p(X_t|X_{t-1} = X)$. Each subpanel includes the corresponding partition of the set of random variables, indicated by variable index sets, similar to figs. \ref{fig:Figure_effect_ex} and \ref{fig:Figure_cause_ex}. Specifically, the indices are  $a_{t+1} = 1, b_{t+1} = 2, c_{t+1} = 3, a_t = 4, b_t = 5, c_t = 6)$. Te particular partition then gives rise to the decomposed variant of the effect repertoire. The asterisk signifies the decomposed effect repertoire which results in the minimum EMD with respect to the original effect repertoire, which in turn corresponds to integrated effect information $\phi_e((1,0,0))$.}\label{fig:Figure_S3}
\end{figure}

\begin{figure}[h!]
\includegraphics*[width=\textwidth]{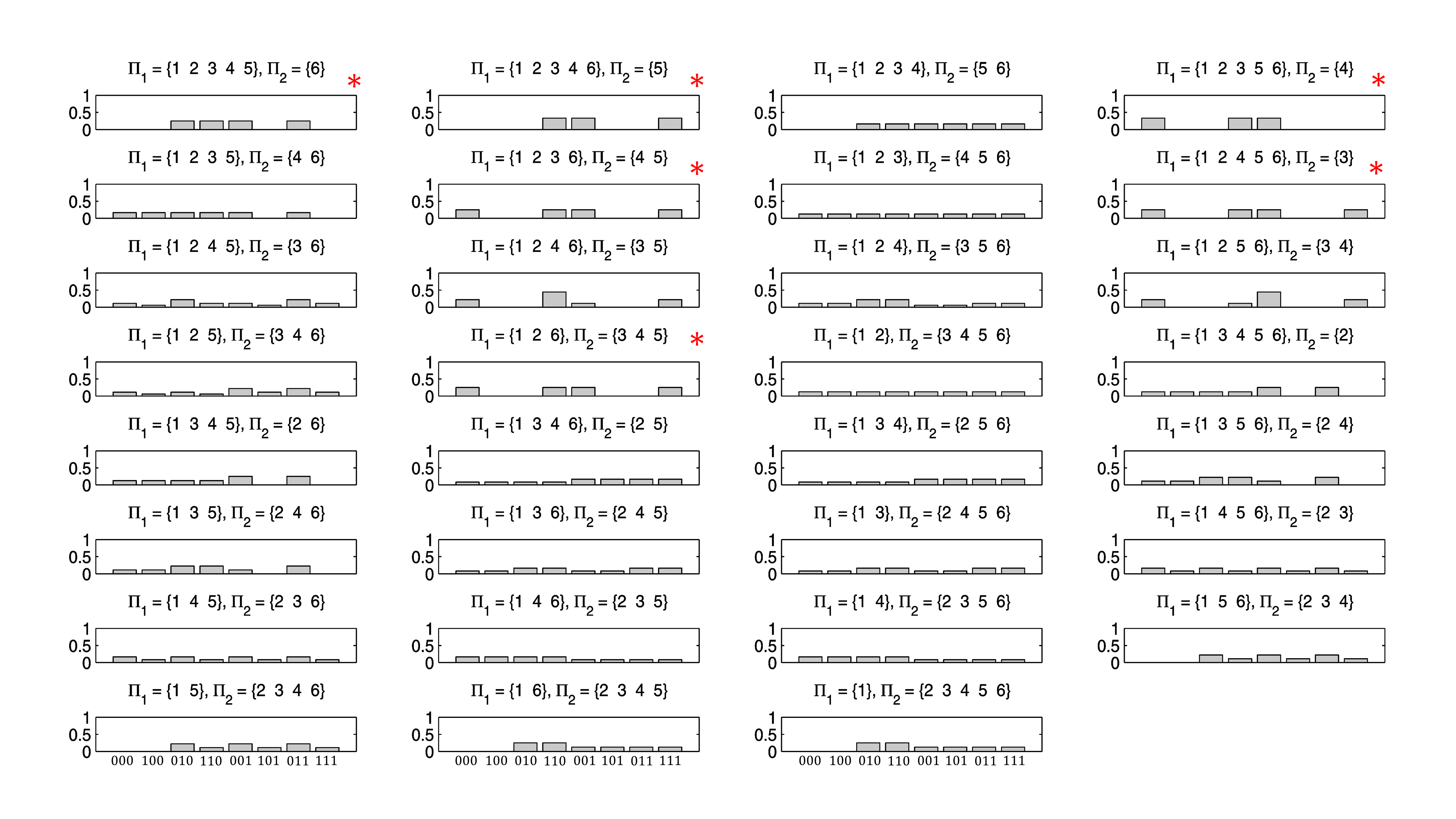}
\caption{{\bf Decomposed cause repertoire for an exemplary discrete-state system.} For the system state $X = (1,0,0)$, the figure visualizes the complete set of decomposed variants of the random variable set $(X_{t-1},X_t)$ giving rise to the decomposed cause repertoire variants. Each subpanel includes the corresponding partition of the set of random variables, indicated by variable index sets, similar to figs. \ref{fig:Figure_effect_ex} and \ref{fig:Figure_cause_ex}. Specifically, the indices are  $a_{t-1} = 1, b_{t-1} = 2, c_{t-1} = 3, a_t = 4, b_t = 5, c_t = 6$. The asterisks signify the decomposed cause repertoires which results in a minimum EMD with respect to the original cause repertoire, which in turn corresponds to integrated cause information $\phi_c((1,0,0))$}\label{fig:Figure_S4}
\end{figure}

As an example of how to compute integrated cause-effect information, Figure \ref{fig:Figure_example_phi} depicts the evaluation of $\phi_{ce}(X)$ for the state $X = (1,0,0)$ in the exemplary discrete state system from the main text. Specifically, the right panels depict the probability mass functions underlying the evaluation of integrated effect information $\phi_e((1,0,0))$, and the left panels the probability mass functions underlying the evaluation the integrated cause information $\phi_c((1,0,0))$. The decomposed variant of the effect repertoire corresponding to partition $i = 16$ (the minimum information partition) results in the minimum EMD $\phi_e((1,0,0)) = 0.25$ with regard to the original effect repertoire. Notably, for the variable subset under scrutiny (the whole system $X^S = \{a, b, c\}$), there are $2^{6-1}-1 = 31$ possible ways to bipartition the graphical model of $p_{ce}(X_{t-1},X_t)$, and hence $31$ versions of $p^{(i)}_e(X_t|X_{t-1})$. For the system state $X = (1,0,0)$, the ensuing distributions over all partitions are shown in Figure \ref{fig:Figure_S3} for the decomposition of the effect repertoire and Figure \ref{fig:Figure_S4} for the decomposition of the cause repertoire. Likewise, the left panel of figure \ref{fig:Figure_example_phi} depicts the cause repertoire of the system state $X = (1,0,0)$. The EMD over the minimum information partition here evaluates to $\phi_c((1,0,0)) = 0.5$. As shown in Figure \ref{fig:Figure_S4}, there are in fact multiple partitions and ensuing distributions, for which this minimum EMD is attained (i.e. the minimum information partition is not necessarily unique in smaller networks, see red asterisks). Evaluating the minimum then results in an integrated cause-effect information of  
\begin{equation}
\phi_{ce}((1,0,0)) = 0.25 
\end{equation}
for the current subset of variables and system state of interest.

\begin{figure}[h!]
\includegraphics*[width=120 mm]{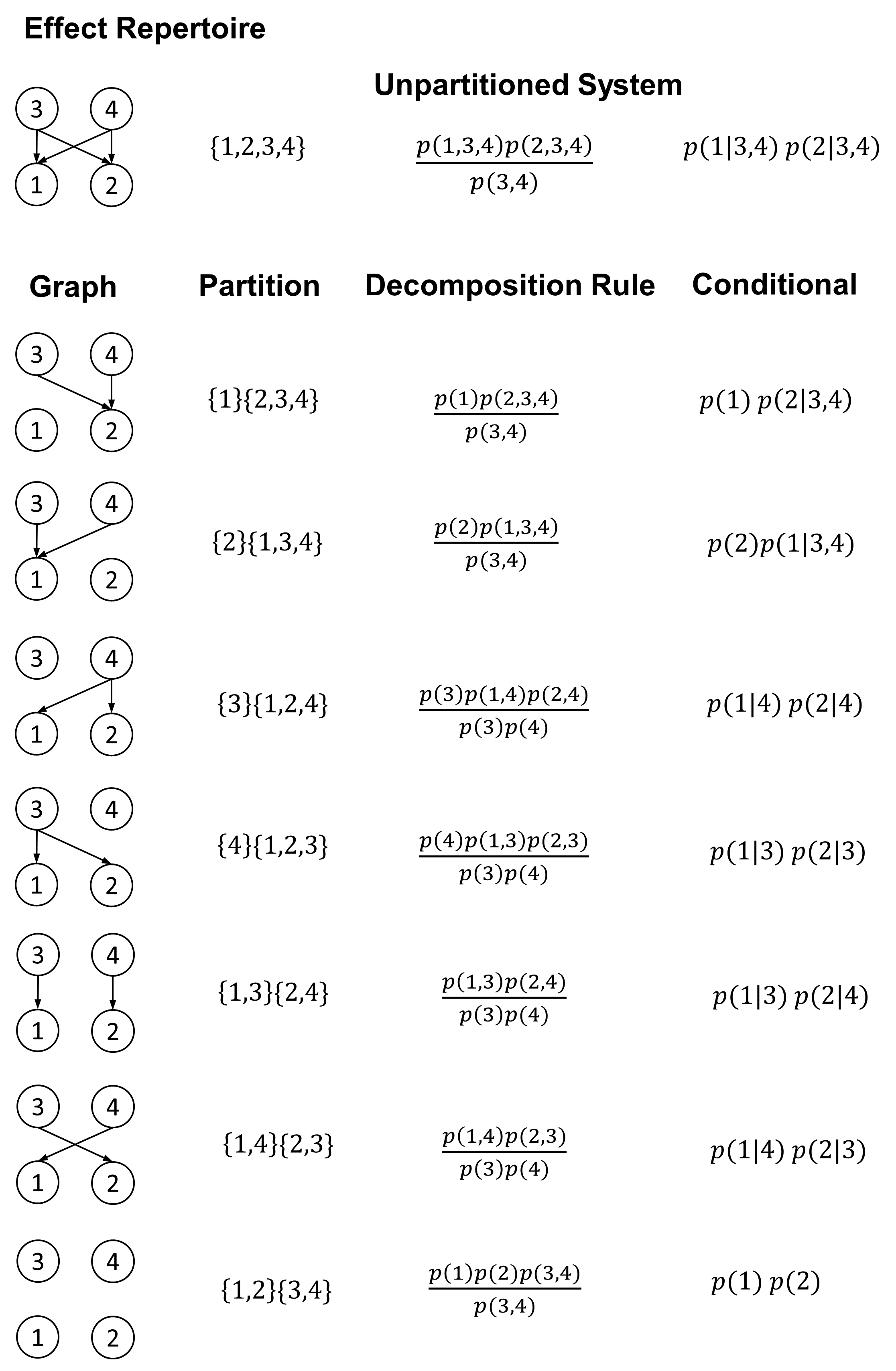}
\caption{{\bf Exhaustive system decomposition (effect repertoire)}. Visualization of all $k=2^{4-1}-1 = 7$ unique bipartitions of a two-dimensional system with $X_t := \lbrace{1,2}\rbrace$ and $X_{t-1} :=\{3,4\}$. For the effect repertoire (first row), the joint distribution over $\lbrace{1,2,3,4}\rbrace$ is factorized according the the formulas in the main text and conditioned on $\lbrace{3,4}\rbrace$. The graphical model for every partition is depicted in the left column alongside the bipartition of $\{1,2,3,4\}$ in the remaining rows. The decomposition rule as presented in the main text is displayed along with the ensuing conditional distribution in the third and fourth columns.}\label{fig:Figure_effect_ex}.
\end{figure}
\newpage
\begin{figure}[htp!]
\includegraphics*[width=120 mm]{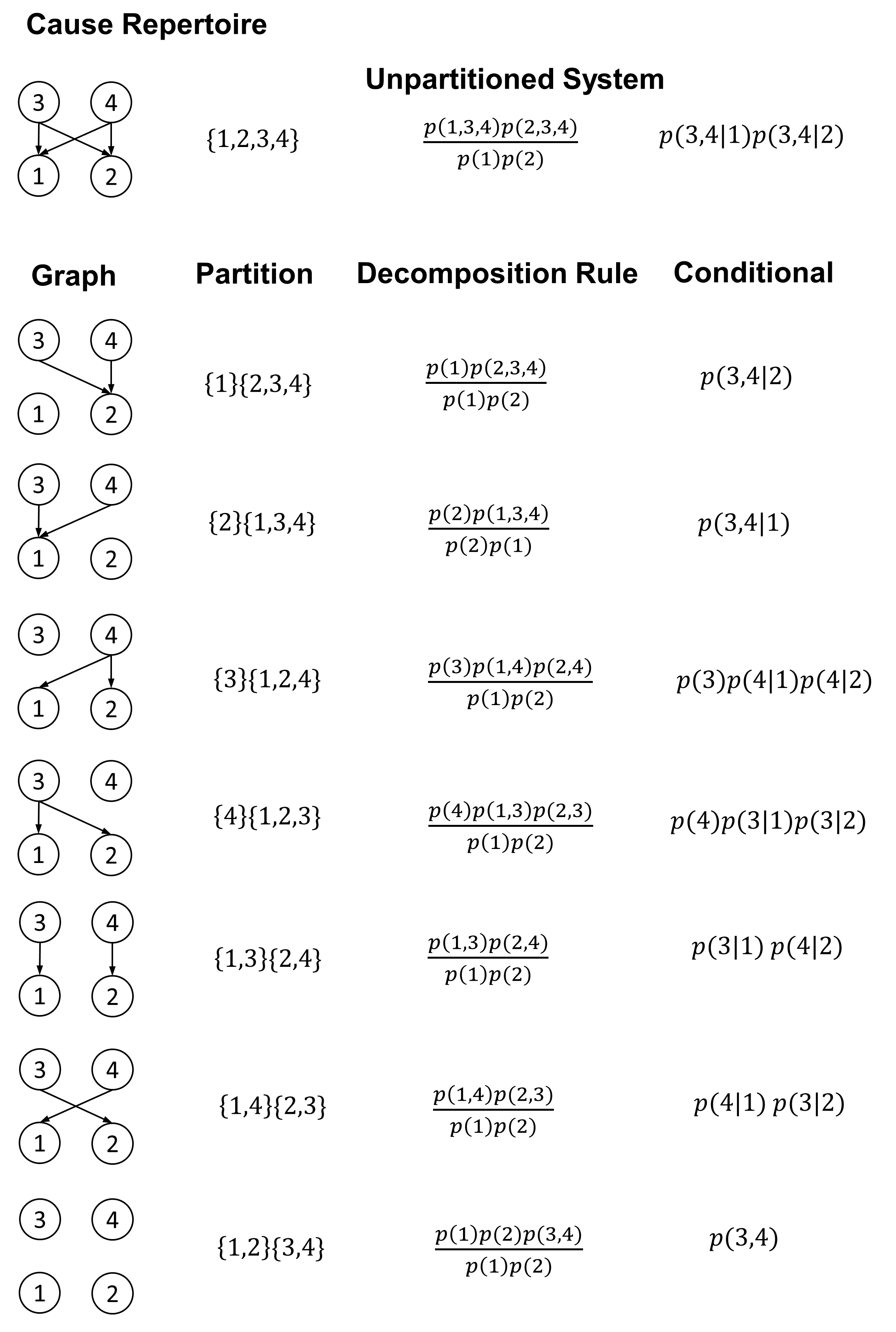}
\caption{{\bf Exhaustive system decomposition (cause repertoire)}. Visualization of all $k=2^{4-1}-1 = 7$ unique bipartitions of a two-dimensional system with $X_t := \lbrace{1,2}\rbrace$ and $X_{t-1} :=\{3,4\}$. For the cause repertoire (first row), the joint distribution over $\lbrace{1,2,3,4}\rbrace$ is factorized according to the formulas from the main text and conditioned on $\lbrace{1,2}\rbrace$. The graphical model for every partition is depicted in the left column alongside the bipartition of $\{1,2,3,4\}$ in the remaining rows. The decomposition rule as presented in the main text is displayed along with the ensuing conditional distribution in the third and fourth columns.}\label{fig:Figure_cause_ex}
\end{figure}

\end{appendix}

\end{document}